\documentclass{aa}
\usepackage{graphicx}
\usepackage{epsfig}
\usepackage{longtable,lscape}
\usepackage{natbib}
\usepackage{rotating}
\usepackage{amssymb}
\bibpunct{(}{)}{;}{a}{}{,} 
\begin{document}

\title{An optical/NIR survey of globular clusters in early-type galaxies}
\subtitle{III. On the colour bimodality of GC systems}

\author{A. L. Chies-Santos\inst{1,2}, S. S. Larsen\inst{2}, M. Cantiello\inst{3}, J. Strader\inst{4}, H. Kuntschner\inst{5}, E. M. Wehner\inst{6} and J. P. Brodie\inst{7}}

\offprints{Ana.Chies\_Santos@nottingham.ac.uk} 

\institute{University of Nottingham, School of Physics and Astronomy, University Park, NG7 2RD Nottingham, UK
\and Astronomical Institute, University of Utrecht, Princetonplein 5, NL-3584, Utrecht, The Netherlands
\and INAF-Osservatorio Astronomico di Teramo, via M. Maggini snc, 64100 Teramo, Italy 
\and Harvard-Smithsonian Center for Astrophysics, Cambridge, MA 02138, USA 
\and European Southern Observatory, Karl-Schwarzschild-Str. 2, 85748 Garching, Germany
\and Departments of Physics and Astronomy, Haverford College, Haverford, PA 19041, USA
\and UCO/Lick Observatory, University of California, Santa Cruz, CA 95064, USA}

\date{Received 2 May 2011/ Accepted 16 January 2012}

\abstract
{ The interpretation that bimodal \textit{colour} distributions of
  globular clusters (GCs) reflect bimodal \textit{metallicity}
  distributions has been challenged.  Non-linearities in the colour to
  metallicity conversions caused for example by the horizontal branch (HB) stars may be
  responsible for transforming a unimodal metallicity distribution
  into a bimodal (optical) colour distribution.}
{We study optical/near-infrared (NIR) colour distributions of the GC
  systems in 14 E/S0 galaxies.
 }
{We test whether the bimodal feature, generally present in optical
  colour distributions, remains in the optical/NIR ones. The latter
  colour combination is a better metallicity proxy than the former. We
  use KMM and GMM tests to quantify the probability that different
  colour distributions are better described by a bimodal, as opposed
  to a unimodal distribution.}
{We find that double-peaked colour distributions are more commonly seen 
in optical than in optical/NIR colours.  For some of
  the galaxies where the optical $(g-z)$ distribution is clearly
  bimodal, a bimodal distribution is not preferred over a unimodal one at a
  statistically significant level for the $(g-K)$ and $(z-K)$ distributions.
  The two most cluster-rich galaxies in
  our sample, NGC\,4486 and NGC\,4649, show some interesting
  differences.   
The $(g-K)$ distribution of NGC\,4649 is better described by a bimodal 
distribution, while this is true for the $(g-K)$ distribution of 
NGC\,4486 GCs only if restricted to a brighter sub-sample with small
K-band errors ($<$ 0.05 mag).
Formally, the K-band photometric errors cannot be responsible for blurring 
bimodal metallicity distributions to unimodal $(g-K)$ colour distributions.
However, simulations including the extra scatter in the colour-colour 
diagrams (not fully accounted for in the photometric errors) show that such 
scatter may contribute to the disappearance of bimodality in $(g-K)$
for the full NGC\,4486 sample.
For the less cluster-rich
galaxies results are inconclusive due to poorer statistics.} 
{A bimodal optical colour distribution is not necessarily an indication of 
an underlying bimodal metallicity distribution. Horizontal branch morphology
may play an important role in shaping some of the optical GC colour
distributions. 
However, we find tentative evidence that the $(g-K)$
colour distributions remain bimodal in the two cluster-rich galaxies in 
our sample (NGC\,4486 and NGC\,4649) when restricted to clusters with
small K-band photometric errors. 
This bimodality becomes less pronounced
when including objects with larger errors, or for the $(z-K)$ colour
distributions. 
Deeper observations of large numbers of GCs will be
required to reach more secure conclusions.}


\titlerunning{On the colour bimodality of GC systems}

\authorrunning{Chies-Santos et al.}

\maketitle

\section{Introduction}
Globular cluster (GC) systems exhibit a bimodal \textit{optical} colour distribution in the majority of luminous and intermediate luminosity early-type galaxies (\citealt{az93}, \citealt{es96}, \citealt{peng06}). This is widely interpreted as being due to the presence of two old sub-populations ($\ga$\,10\,Gyrs) that differ in metallicity (\cite{bs06} and references therein). Metallicity bimodality is obvious for the Milky Way (\citealt{zinn85}, \citealt{bica06}) but seems less evident for our spiral neighbour M31 (e.g. \citealt{caldwell11}). 

The presence of two peaks in the GC colour distributions has been often taken as an argument in favour of the existence of two major epochs/mechanisms of star formation in the host galaxies. There has been much debate in the literature over the past decades as to what is responsible for this bimodality (see \citealt{bs06}). 
Among the scenarios for GC formation that account for it, one can list \cite{az92}, \cite{fbg97}, \cite{cote98}, \cite{beasley02}, \cite{strader05} and \cite{rhode05}. 
All of the above assumed different formation channels for blue (metal-poor) and red (metal-rich) clusters.
However, the properties of these cluster populations are generally not too distinct. 
For instance, \cite{peng08} studying the specific frequencies of 100 early-type galaxies from the Virgo Cluster Survey find basically the same trends when separating the GC systems in red and blue. \cite{peng06} analysing the GC systems colour distributions of the same data set find that there is great similarity between the GC peak-metallicity galaxy-mass relations for the two populations. This implies that the conditions of GC formation for metal-poor and metal-rich GCs could not have been too different.  

The complexity of the formation histories of early-type galaxies, within the hierarchical merging framework (e.g. \citealt{renzini06}), are not expected to naturally produce the near universality of bimodal GC metallicity distributions. This is actually the case for models of GC formation built upon this hierarchical framework (e.g. \citealt{beasley02}) where bimodality occurs only after  introducing a mechanism that artificially truncates the formation of metal-poor GCs. Nonetheless, more recently, \cite{muratov10} introduce a scenario in which the formation of the two metallicity sub-populations of GCs may be a natural outcome of the hierarchical theory of galaxy formation in some, although not the entire range of model realisations. 
 \cite{muratov10} prescribe the formation of GCs semi-analytically using assembly histories from cosmological simulations combined with observed scaling relations for the amount of cold gas available for star formation.

The interpretation that colour distributions translate directly into metallicity distributions was challenged by \cite{yoon06}. They demonstrate that non-linear colour-metallicity relations caused by the horizontal branch morphology (HB) may transform a \textit{unimodal metallicity distribution} into a \textit{bimodal optical colour distribution}. This issue has been investigated in more detail by \cite{cb07}, who conclude that combinations of optical and near-infrared (NIR) colours are much less sensitive to this effect. \cite{richtler06} had already shown that a flat metallicity distribution can result in a bimodal colour distribution using the Washington photometric system (acknowledging discussions with Boris Dirsch).

A few observational attempts to address whether optical 
colour bimodality is really representative of metallicity bimodality
  exist in the literature, providing sometimes ambiguous or
  conflicting results. For example, \cite{strader07} find clear
  evidence for two metallicity subpopulations in the spectroscopic
  sample of 47 NGC\,4472 GCs from \cite{cohen03}. For the same set of
  data, though, the analysis of \citeauthor{cohen03} do strongly
  favour bimodality. \cite{kundu07} find an optical/NIR $(I-H)$
  bimodal distribution for NGC\,4486 in a small sample of GCs from one
  NICMOS/HST pointing combined with HST/WFPC2 data. 
  Moreover, \cite{spitler08} presented an optical/mid--infrared analysis using
  Spitzer Space Telescope for NGC\,4594 and NGC\,5128.  Both galaxies
  present a clear optical bimodality. While NGC\,5128 presents an
  obvious R-[3.6] bimodal distribution, more compatible with the
  multimodal peaks that the spectroscopic work of \cite{beasley08}
  find (see also \citealt{woodley10}), NGC\,4594 has a less clear
  bimodal distribution in this baseline. In contrast, \cite{alves11} present spectra for over 200 GCs in the Sa NGC\,4594
and find a clear bimodal distribution. More recently,
\cite{foster10} and \cite{foster11} find very similar Calcium triplet values for red and
blue GCs in NGC\,1407 and NGC\,4494 despite their large colour difference. Since
Calcium triplet is a metallicity indicator, similar values for red and
blue GCs indicate similar metallicities. One possible explanation
given by the authors is the non-linear conversion between colour and
metallicity. 
  
 \cite{blakeslee10} simulate GC populations with both a
mass-metallicity relation (\textit{blue-tilt}) and a non-linear
colour-metallicity relation and find bimodal colour distributions with
a \textit{blue-tilt} even though the metallicity distribution appears
unimodal.

Finally, the very recent works by \cite{yoon11a} and
  \cite{yoon11b} show that colours such as $(u-g)$ and $(u-z)$ have
  significantly less inflected colour metallicity relations than
  $(g-z)$. They also show that the metallicity distributions obtained
  from inflected colour metallicity relations are strongly-peaked,
  unimodal and with a broad metal-poor tail, similar to that of the
  resolved field stars in nearby elliptical galaxies and those
  produced by chemical evolution models of galaxies (e.g. \citealt{bird10}).

The purpose of this work is to investigate the nature of
  optical/NIR colour distributions of different GC systems in several
  early-type galaxies. We aim to shed light on the true nature of the
  metallicity distributions of GC systems.

\section{Observations and data}
The observations and data reduction techniques of the data used in this study are described thoroughly in \cite{paper1}. Here we briefly summarise the procedures applied.
A sample of 14 early-type galaxies was imaged in the Ks-band (from now on referred to as only $K$) with LIRIS at the WHT and combined with archival ACS/HST F475W ($\sim$ $g$) and F814LP ($\sim$ $z$) images.
The galaxies have $M_B\,<\,-19$ and $(m-M)\,< \,32$.
ACS images were reduced with MULTIDRIZZLE (\citealt{koek02}) and LIRIS images with LIRISDR, in addition to standard IRAF routines.

The GCs were detected and had sizes ($R_{eff}$) measured in the ACS images with DAOFIND and ISHAPE (\citealt{larsen99}), respectively.
Aperture photometry was performed in the $g$, $z$ and $K$ bands with PHOT (\citealt{stetson87}). After the following criteria were applied to automatically detected sources, $g<23$, $0.5<(g-z)<2.0$, $1<R_{\rm eff}({\rm pc})<15$ a careful visual inspection was performed where obvious non-cluster objects were removed.
Also, sources that were too close together in the ACS images and that appeared as one bigger source in the LIRIS images, due to its lower resolution compared to ACS, were flagged. 
Finally, we caution the reader that the data of NGC\,4382 and NGC\,4473 were taken at non-photometric conditions.

\section{Integrated colours and horizontal branch morphology}
In this section we study the effect of the HB in the integrated colours of GCs.
The colour of the HB varies abruptly between [Fe/H]$=-0.6$ and $-0.9$. At this [Fe/H] range the HB departs from the red-clump position (\citealt{lee94}). 
This is argued as the cause of the non-linear behaviour of the colour-metallicity relation (\citealt{yoon06}).
\cite{peng06} with data from the Milky Way, NGC\,4486 and NGC\,4472 showed that the empirical transformation from metallicity to the optical $(g-z)$ colour is clearly non-linear. 
\cite{d03} had already called attention to the fact that the colour-metallicity relation is non-linear and that starting from the bluest colours equidistant colour intervals are projected onto 
progressively larger metallicity intervals. 

\begin{figure}
\begin{center}
\includegraphics[width=7cm]{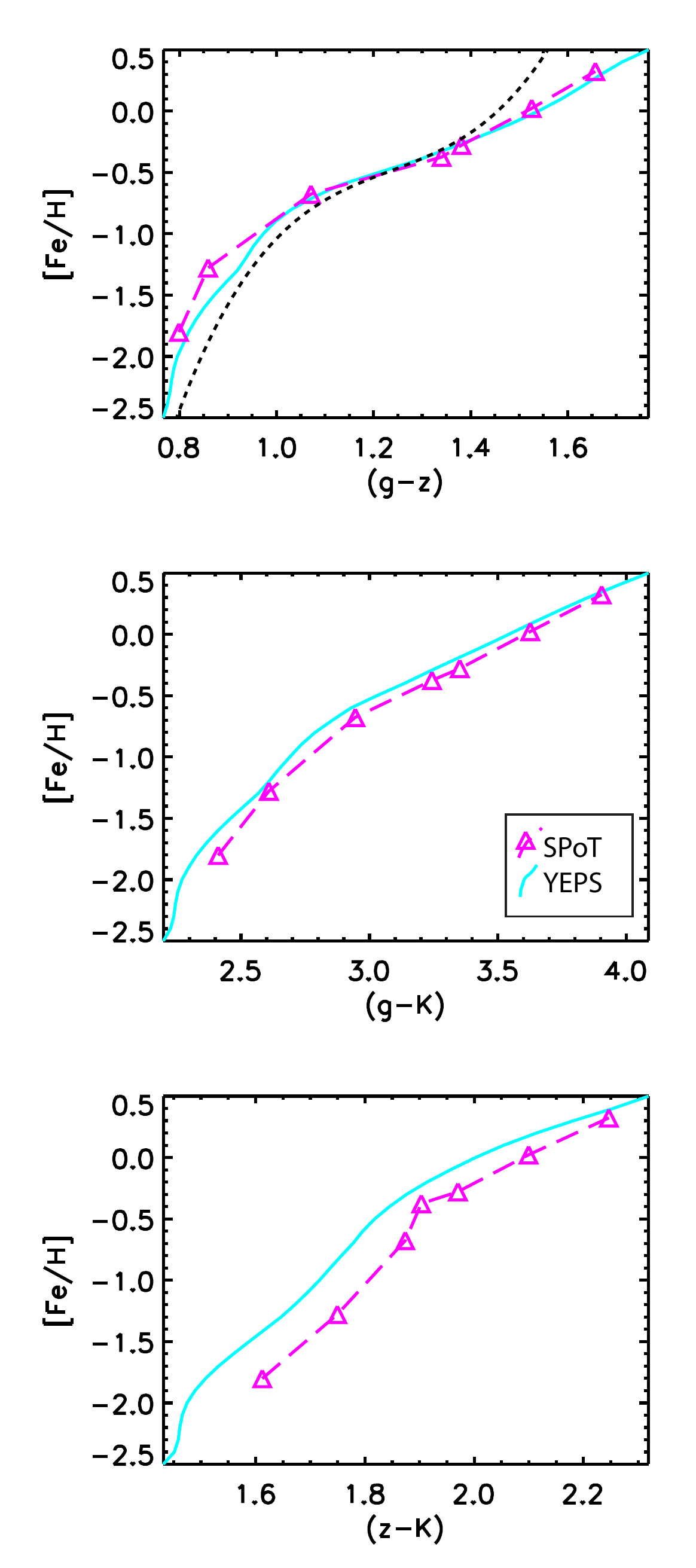}
 \caption{14$\,$Gyr SPoT and YEPS $(g-z)$, $(g-K)$ and $(z-K)$ colour-metallicity relations according to the legend in the middle panel. The black short-dashed-line in the first panel corresponds to the empirical relation to the \cite{peng06} data for Milky Way and Virgo Cluster GCs  (\citealt{blakeslee10}).}
 \label{col_met}
\end{center}
\end{figure}

In Fig. \ref{col_met} $(g-z)$, $(g-K)$ and $(z-K)$ colour-metallicity
relations are shown. For those relations we used 14$\,$Gyr
SPoT\footnote{SPoT models can be downloaded from
  www.oa-teramo.inaf.it/spot.} (\citealt{raimondo05},
\citealt{biscardi08}) and YEPS\footnote{YEPS models can be
    downloaded from http://web.yonsei.ac.kr/cosmic/data/YEPS.htm}
  models. The models for $(g-z)$ show a clear departure from
linearity around [Fe/H]$\sim-0.5$; the wiggly feature noticed by
\cite{yoon06}.  This wiggle that is also visible in other commonly
used colours such as $(V-I)$ and $(B-I)$ (\citealt{cb07}) is generated
by the transition from blue to red HBs. This feature is possibly also
recognizable in $(g-K)$, although very mildly. Such feature is not
visible when considering $(z-K)$.  This last colour does however show
a non-linearity at intermediate metallicities more
pronounced for the SPoT models with respect to the YEPS. This is
probably due to an interpolation effect as the point responsible for
this, [Fe/H] $\sim -0.4$, is right at the transition between blue and
red HBs. Nevertheless the sign of the second derivative of the
$(z-K)\,-\,$[Fe/H] relation at the non-linearity is opposite from the
one for $(g-z)\,-\,$[Fe/H]. This indicates the different nature of the
non-linearities in these 2 colour spaces. If one would remove the
[Fe/H] $\sim -0.4$ point, the $(z-K)$ model would bear linearity.
The SPoT and YEPS models agree well for the $(g-z)$ and
  $(g-K)$ colour$\,-\,$metallicity relations. For the $(z-K)$
  colour$\,-\,$metallicity relation there is more disagreement between
  the models, with the YEPS models being systematically bluer in this
  colour than SPOT for a given metallicity. This also happens for
  $(g-K)$ but to a smaller extent.

\begin{figure}
\begin{center}
\includegraphics[width=7cm]{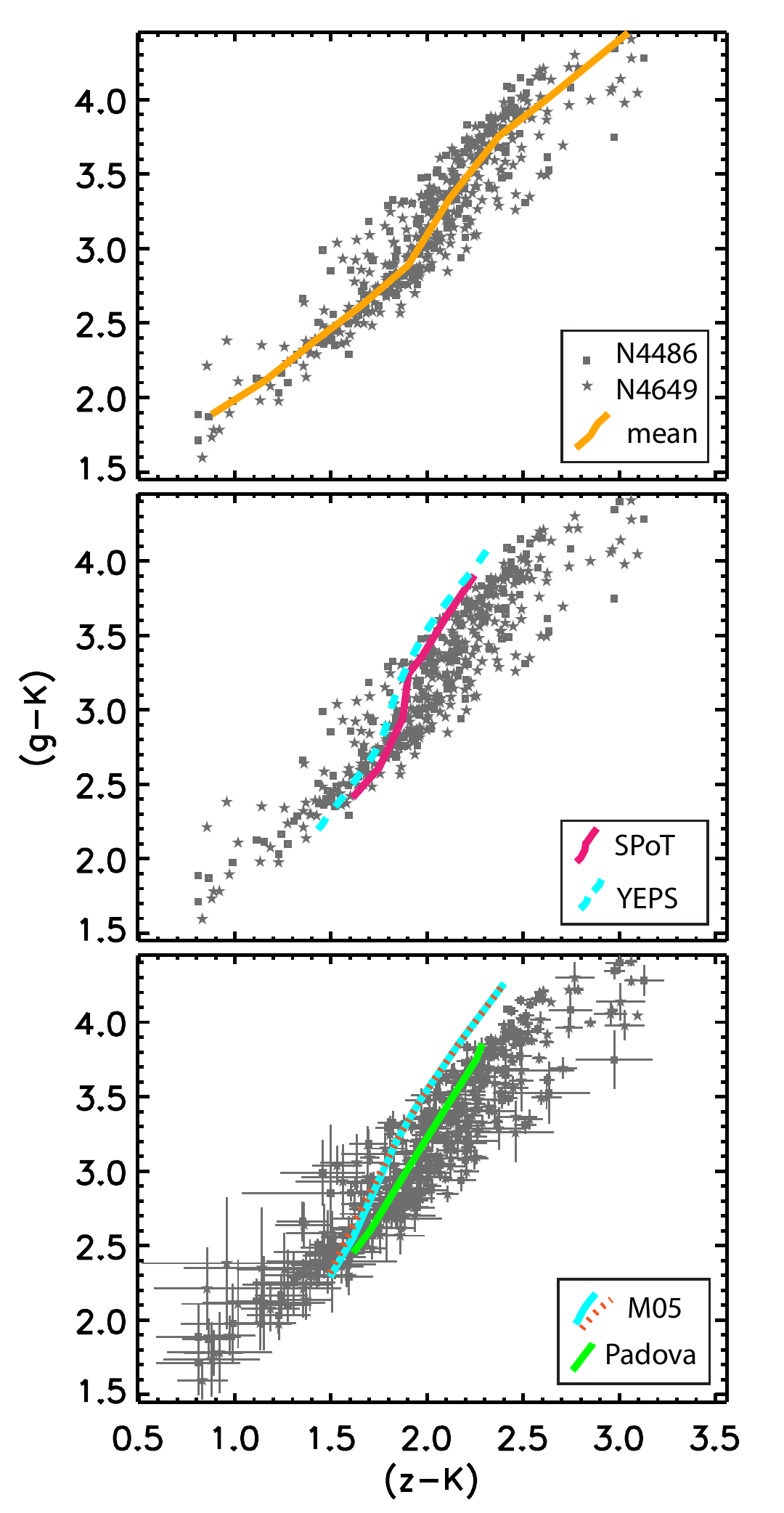}
 \caption{$(z-K)$ \textit{vs.} $(g-K)$ for GCs of NGC\,4486 and
   NGC\,4649, indicated by different symbols and with error bars in
   the bottom panel.  Note the wavy feature the data presents around
   $(z-K)\,\sim\,2$ and $(g-K)\,\sim\,3.2$. \textit{Upper panel}: a
   running median on the data points; \textit{Middle panel}: 14\,Gyr
   SpOT-Teramo and YEPS SSPs with a realistic treatment of
   the HB morphology (\citealt{raimondo05}, \citealt{yoon06},
   \citealt{yoon11a}, \citealt{yoon11b}); \textit{Bottom panel}:
   \cite{m05} with blue and red HB at the lowest metallicities and
   Padova08 14 Gyr SSPs. See Fig.\,16 of \cite{paper2} for a $(g-K)$
   \textit{vs.} $(g-z)$ plot.}
 \label{wiggle}
\end{center}
\end{figure}   

In Fig. \ref{wiggle} the $(g-K)-(z-K)$ diagram for the joint NGC\,4486
and NGC\,4649 GC sample is shown with different simple
  stellar population (SSP) models over plotted in different
panels. Note the wavy feature the data presents around
$(g-K)\,\sim\,3.2$ and $(z-K)\,\sim\,2$. A running median on the data
points is over plotted in the top panel for better visualisation of
the wiggly behaviour. The SPoT 14 Gyr SSP (middle panel) is over
plotted where this non-linear feature is also present.  SPoT models
match the observed colour magnitude diagrams of star clusters fairly
well over a wide range of ages and chemical compositions
(\citealt{brocato00}, \citealt{raimondo05}). In fact, SPoT models are
optimised to simulate the HB spread observed in galactic GCs and this
may be one point of uncertainty. \cite{sohn06} has shown evidence that
the most massive NGC\,4486 GCs may have different He abundances and
thus different HB morphology than galactic GCs. The YEPS
  models are also shown in Fig. \ref{wiggle}. As SPoT, the YEPS
models consider the systematic variation in the mean colours of HB
stars as a function of metallicity.  Even though the SPoT and
  YEPS tracks in Fig. \ref{wiggle} do not fit the data well in the
redder part of the diagram, they present a wavy feature approximately
at the same location as the data: the location pinpointing the
transition from blue to red HB morphology.  Other SSP models, in which
the detailed modelling of the metallicity dependent HB morphology of
the SPoT/YEPS models is not included, do not show this
behaviour (bottom panels for Padova08\footnote{Padova SSPs retrieved
  from the CMD 2.2 input form (http://stev.oapd.inaf.it/cmd), with
  \cite{marigo08} isochrones} and for \citealt{m05}).  The
SPoT/YEPS models for the other 2-colour combinations also
present this non-linear behavior, although it is not possible to see
it in the data using $(g-K)$ \textit{vs.}  $(g-z)$ and $(g-z)$
\textit{vs.} $(z-K)$ as clearly as it can be seen using $(g-K)$
\textit{vs.} $(z-K)$.  We suspect that this could be an effect of
scatter as the $K_{err}$ is dominant.  The errors in $(g-K)$ and
$(z-K)$ are not actually independent, but both dominated by the
$K_{err}$. This means that the uncertainties cause the data to scatter
more or less diagonally in the plot.  Moreover this diagram has a more
similar scale than with the other combinations, where only one of the
colours contains $K$.  We refer the reader to Fig. 16 of \cite{paper2}
for a plot of $(g-K)$ \textit{vs.} $(g-z)$.

By excluding the [Fe/H] $\sim -0.4$ point for the SPoT
  models, the $(z-K)$--[Fe/H] would be more linear.  However, a
  similar behaviour is found also for the YEPS models at approximately
  this metallicity. 
  The $(g-K)$
  colour presents a near linear relation to metallicity and also has a
  broader baseline than $(z-K)$.

\section{Colour distributions and bimodality tests}
\begin{figure*}[h]
\centering
 \hspace*{1.9cm}
\includegraphics[width=150mm]{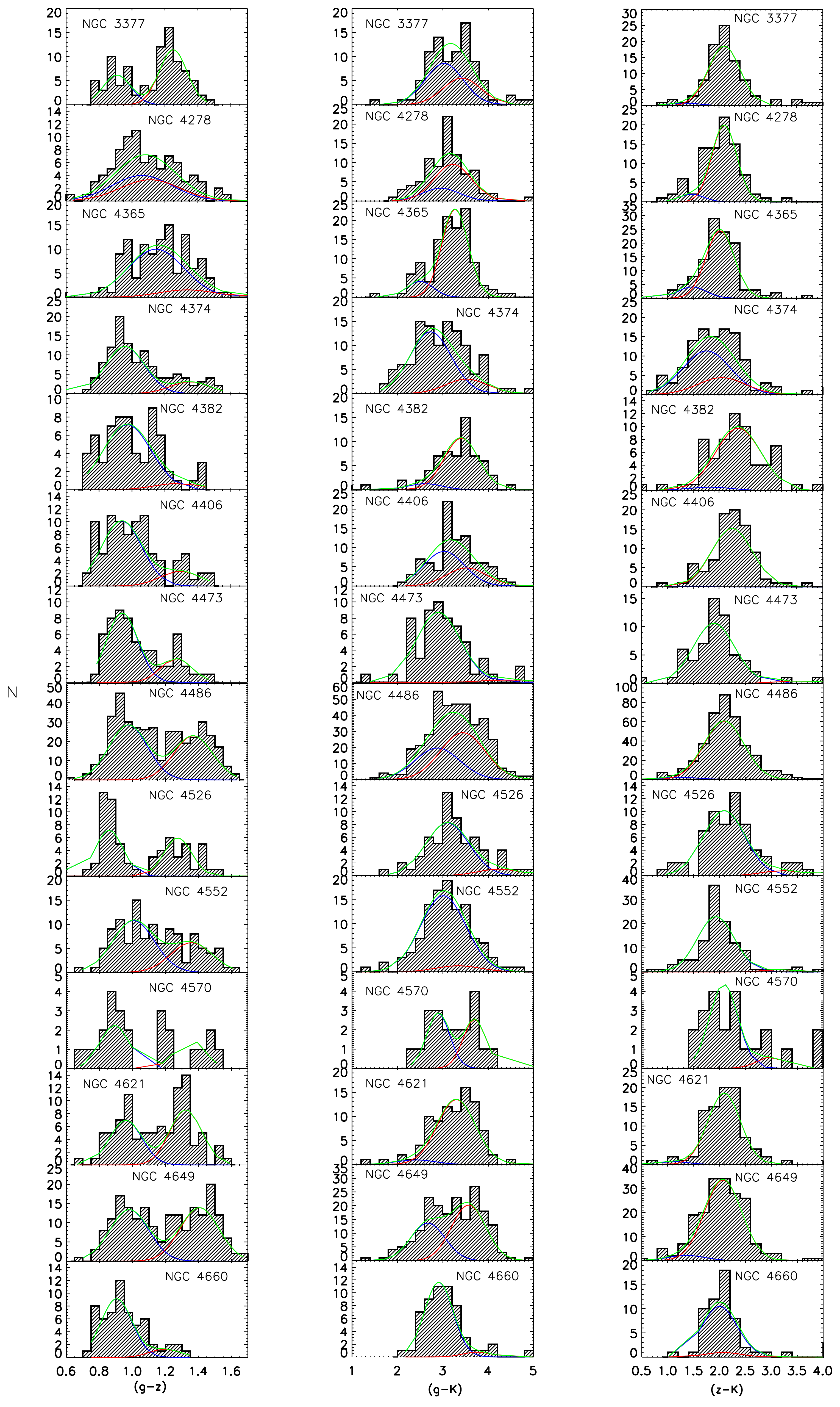}
\caption{$(g-z)$, $(g-K)$ and$(z-K)$ colour distribution for the galaxies. Over plotted are the bimodal (blue and red peaks in the corresponding colours) distributions returned by KMM and the sum of the blue and red peaks (in green).}
\label{kmmfigs}
\end{figure*}

In this section we investigate the behaviour of the $(g-z)$, $(g-K)$ and $(z-K)$ colour distributions. They are shown for the different GC systems in Fig. \ref{kmmfigs}. Note that while a great number of systems look bimodal in $(g-z)$ they appear less and less bimodal as one moves on to the colours that should in principle sample less the horizontal branch ($(g-K)$ and $(z-K)$). The bimodality feature is generally the least apparent in the $(z-K)$ histograms.
To quantify whether the distributions are better described by bimodal as opposed to unimodal models, the KMM \citep{abz94} algorithm was applied to the data. 
The bimodal (blue and red peaks and their sum) distribution estimates returned by the code are shown in
Fig. \ref{kmmfigs}. The outputs of KMM for the different colours and galaxies are listed in Table \ref{kmm}.
This test utilises the likelihood ratio test (LRT) to estimate the probability that the distribution of a number of data values is better modelled as a sum of two Gaussian distributions than a single Gaussian. 
However, it obeys a $\chi^2$ statistics only when the two modes have the same variance (\textit{homoscedastic} case). P(KMM) indicates the probability of rejecting a unimodal distribution in favour of a bimodal one, with a low probability indicating that a bimodal distribution is preferred.

\begin{table}
\begin{scriptsize}
\centering
\begin{tabular}{cccccccccccccc}

\hline
(1)         & (2)            & (3)       &             (4)       &      (5)         &    (6)       &        (7)                    \\
\hline
     Galaxy   & $N_{b}$ & $N_{r}$    & $\mu_{b}$   & $\mu_{r}$ & $\sigma$ & $P(KMM)$  \\
\hline                                                                               
$n3377_{(g-z)}$ & 26  &    48        &   0.909      & 1.25      &  0.084   &   0.000  \\
$n3377_{(g-k)}$ & 45  &    29        &   3.03       & 3.42      &  0.419   &   0.974  \\ 
$n3377_{(z-k)}$ & 3   &    71        &   1.33       & 2.10      &  0.306   &   0.791  \\
\hline                                                                               
$n4278_{(g-z)}$ & 36  &    30        &   1.06       & 1.11      &  0.181   &   1.000  \\
$n4278_{(g-k)}$ & 17  &    49        &   2.96       & 3.22      &  0.411   &   0.998  \\
$n4278_{(z-k)}$ & 6   &    60        &   1.47       & 2.10      &  0.241   &   0.387  \\
\hline                                                                               
$n4365_{(g-z)}$ & 85  &    13        &   1.15       & 1.33      &  0.170   &   0.979 \\
$n4365_{(g-k)}$ & 15  &    83        &   2.50       & 3.27      &  0.292   &   0.029 \\  
$n4365_{(z-k)}$ & 14  &    84        &   1.48       & 2.02      &  0.274   &   0.867 \\
\hline
$n4374_{(g-z)}$ & 72  &    18        &   0.958      & 1.35      &  0.119   &   0.000 \\ 
$n4374_{(g-k)}$ & 73  &    17        &   2.73       & 3.52      &  0.455   &   0.513 \\  
$n4374_{(z-k)}$ & 65  &    25        &   1.76       & 2.05      &  0.459   &   0.999 \\  
\hline                                                                               
$n4382_{(g-z)}$ & 52  &    5         &   0.971      & 1.28      &  0.145   &   0.350 \\
$n4382_{(g-k)}$ &  6  &    51        &   2.61       & 3.41      &  0.381   &   0.352 \\
$n4382_{(z-k)}$ &  3  &    54        &   1.82       & 2.37      &  0.443   &   0.862 \\
\hline                                                                               
$n4406_{(g-z)}$ & 62  &   14         &   0.94       & 1.29      &  0.123   &   0.006 \\
$n4406_{(g-k)}$ & 50  &   26         &   3.03       & 3.52      &  0.443   &   0.949 \\  
$n4406_{(z-k)}$ &  1  &   75         &   1.39       & 2.25      &  0.395   &   0.692 \\
\hline                                                                               
$n4473_{(g-z)}$ &  41  &   14        &   0.94       & 1.27      & 0.097    &   0.001 \\
$n4473_{(g-k)}$ &  53  &   2         &   2.90       & 4.42      & 0.486    &   0.055 \\  
$n4473_{(z-k)}$ &  53  &   2         &   1.90       & 3.56      & 0.400    &   0.007 \\
\hline
$n4486_{(g-z)}$ & 167  &  134        &   0.97       & 1.37      & 0.117   &   0.000 \\
$n4486_{(g-k)}$ & 121  &  180        &   2.88       & 3.44      & 0.493   &   0.934 \\  
$n4486_{(z-k)}$ & 10   &  291        &   1.24       & 2.09      & 0.383   &   0.122 \\
\hline
$n4526_{(g-z)}$ & 30  &	  25         &   0.859      & 1.27      & 0.084   &   0.000 \\
$n4526_{(g-k)}$ & 49  &    6         &   3.10       & 4.24      & 0.475   &   0.669 \\  
$n4526_{(z-k)}$ & 51  &    4         &   2.09       & 3.22      & 0.402   &   0.114 \\
\hline
$n4552_{(g-z)}$ & 68  &  39          &   1.01       & 1.36      & 0.126   &   0.003 \\
$n4552_{(g-k)}$ & 99  &  8           &   3.01       & 3.34      & 0.499   &   0.999 \\  
$n4552_{(z-k)}$ & 103 &  4           &   1.93       & 3.19      & 0.358   &   0.007 \\
\hline
$n4570_{(g-z)}$ & 10  &  7           &   0.89       & 1.35      & 0.089   &   0.004 \\
$n4570_{(g-k)}$ & 9   &  8           &   2.90       & 3.70      & 0.251   &   0.196 \\  
$n4570_{(z-k)}$ & 15  &  2           &   2.08       & 3.00      & 0.274   &   0.180 \\
\hline
$n4621_{(g-z)}$ & 34  & 42           &   0.96       & 1.32      & 0.098   &  0.000 \\
$n4621_{(g-k)}$ &  5  & 71           &   2.41       & 3.30      & 0.421   &  0.284 \\  
$n4621_{(z-k)}$ &  3  & 73           &   1.08       & 2.09      & 0.315   &  0.040 \\
\hline
$n4649_{(g-z)}$ &  79 & 82           &   0.98       & 1.41      & 0.118   &  0.000 \\
$n4649_{(g-k)}$ &  65 & 96           &   2.67       & 3.57      & 0.379   &  0.020 \\  
$n4649_{(z-k)}$ &  10 & 151          &   1.36       & 2.07      & 0.360   &  0.173 \\
\hline
$n4660_{(g-z)}$ &  43  & 6           &   0.90       & 1.19      & 0.094   &  0.021 \\
$n4660_{(g-k)}$ &  46  & 3           &   2.91       & 3.76      & 0.316   &  0.194 \\  
$n4660_{(z-k)}$ &  45  & 4           &   2.00       & 2.07      & 0.342   &  1.000 \\

\hline

\end{tabular}
\caption{KMM outputs for the colour distributions of the galaxies: (1) galaxy and its colour distributions, (2) number of clusters assigned to the blue peak ($N_{b}$), (3) number of clusters assigned to the red peak ($N_{r}$), (4) the mean of the blue peak ($\mu_{b}$), (5) the mean of the red peak ($\mu_{r}$), (6) the common width of the peaks ($\sigma$) and (7) the probability for rejecting a unimodal distribution (P(KMM)).}
\label{kmm}
\end {scriptsize}
\end{table}

The GMM (\citealt{muratov10}) code was also applied to the data.  GMM
is a robust generalisation of the KMM test. Besides modelling a
bimodal distribution with different mode variances
(\textit{heteroscedastic} case), it uses three different statistics:
LRT($\chi^{2}$), distance from the mean peaks (\textit{DD}) and
kurtosis (\textit{kurt}). Moreover, it makes use of bootstrapping to
derive uncertainties for the estimators of these different
statistics. Finally, GMM outputs the probability that the three
statistics give for rejecting a unimodal fit in favour of a bimodal
one: P($\chi^{2}$), P(DD) and P(kurt). From now on, we refer to
P($\chi^{2}$) as P(GMM).  The \textit{DD} statistic is a measure of
how meaningful the split between the two peaks is.  The kurt statistic
is a measure of \textit{peakedness} of the distribution. A positive
kurtosis corresponds to a sharply peaked distribution whereas a
negative kurtosis corresponds to a flattened distribution. A negative
kurtosis is a necessary but not sufficient condition for
bimodality. The \textit{DD} and \textit{kurt} are additional checks to
support the results of the LRT since the likelihood function is very
sensitive to outliers far from the centre of the distribution and may
reject a genuine unimodal distribution.  

Before running KMM and GMM objects out of the ranges
  $1<(g-K)<5$ and $0.5<(z-K)<4$ were removed from the
  distributions. These are objects far from the centre of the
  distributions and are very likely to affect the results of the
  bimodality tests by e.g. creating artificial peaks which will be
  interpreted as separate populations by these statistical tests. For
  NGC\,4486, there is no object in this range, while for NGC\,4649
  three objects drop off the distribution with these restrictions.
The outputs of GMM for the different colours and galaxies are listed
in Table \ref{gmm}.  For the galaxy NGC\,4570, due to the low number
of clusters, the test did not converge, and these estimates are not
shown.

\begin{landscape}
\begin{table}
\begin{scriptsize}
\begin{tabular}{cccccccccccccccccc}
\hline
             (1)  & (2) & (3)   & (4)     &      (5)   &  (6)  &  (7) & (8) & (9) & (10) & (11) & (12)\\
\hline
      Galaxy & $N_{b}$ & $N_{r}$ & $\mu$ & $\mu_{b}$  &$\mu_{r}$ & $\sigma$ & $\sigma_{b}$ &	$\sigma_{b}$ & P(GMM) & P(DD) & P(kurt)  \\
\hline                                                                               
$n3377_{(g-z)}$ & 24.7$\pm$4.7 & 49.3$\pm$4.70 & 1.130$\pm$0.021 & 0.899$\pm$0.018 & 1.247$\pm$0.016 & 0.183$\pm$0.011 & 0.065$\pm$0.013 & 0.090$\pm$0.011 & $<$ 0.001 & 0.055 &0.001\\
$n3377_{(g-k)}$ & 27.1$\pm$16.7 & 46.9$\pm$16.7 & 3.194$\pm$0.054 & 2.798$\pm$0.226 & 3.455$\pm$0.291 & 0.456$\pm$0.042 & 0.236$\pm$0.119 & 0.349$\pm$0.118 & 0.474 & 0.763 & 0.643 \\
$n3377_{(z-k)}$ & 29.2$\pm$19.4 & 44.8$\pm$19.4 & 2.063$\pm$0.042 & 1.805$\pm$0.367 & 2.155$\pm$0.209 & 0.340$\pm$0.037 & 0.286$\pm$0.175 & 0.270$\pm$0.123 & 0.371 & 0.966 & 0.966 \\
\hline
$n4278_{(g-z)}$ & 40.4$\pm$16.8 & 25.6$\pm$16.8 & 1.085$\pm$0.020 & 0.981$\pm$0.085 & 1.265$\pm$0.101 & 0.178$\pm$0.014 & 0.113$\pm$0.039 & 0.115$\pm$0.051 & 0.653 & 0.530 & 0.38 \\
$n4278_{(g-k)}$ & 30.3$\pm$20.7 & 35.7$\pm$20.7 & 3.101$\pm$0.051 & 2.801$\pm$0.344 & 3.320$\pm$0.283 & 0.425$\pm$0.038 & 0.287$\pm$0.148 & 0.308$\pm$0.135 & 0.601 & 0.952 & 0.666 \\
$n4278_{(z-k)}$ & 20.0$\pm$16.6 & 46.0$\pm$16.6 & 2.016$\pm$0.039 & 1.645$\pm$0.312 & 2.105$\pm$0.117 & 0.308$\pm$0.030 & 0.214$\pm$0.147 & 0.208$\pm$0.074 & 0.318 & 0.857 & 0.878 \\
\hline
$n4365_{(g-z)}$ & 24.4$\pm$19.1 & 73.6$\pm$19.1 & 1.183$\pm$0.020 & 0.965$\pm$0.065 & 1.264$\pm$0.105 & 0.185$\pm$0.013 & 0.063$\pm$0.034 & 0.152$\pm$0.034 & 0.086 & 0.503 & 0.731 \\
$n4365_{(g-k)}$ & 18.8$\pm$14.3 & 79.2$\pm$14.3 & 3.129$\pm$0.045 & 2.466$\pm$0.232 & 3.256$\pm$0.089 & 0.417$\pm$0.032 & 0.190$\pm$0.124 & 0.299$\pm$0.045 & 0.022 & 0.120 & 0.553 \\
$n4365_{(z-k)}$ & 48.6$\pm$17.2 & 49.4$\pm$17.2 & 1.945$\pm$0.034 & 1.861$\pm$0.157 & 2.059$\pm$0.224 & 0.332$\pm$0.033 & 0.318$\pm$0.127 & 0.242$\pm$0.129 & 0.051 & 0.862 & 0.948 \\
\hline
$n4374_{(g-z)}$ & 72.7$\pm$9.1  & 17.3$\pm$9.1  & 1.038$\pm$0.021 & 0.964$\pm$0.025 & 1.369$\pm$0.073 & 0.197$\pm$0.016 & 0.125$\pm$0.026 & 0.093$\pm$0.036 & $<$0.001 & 0.120& 0.756\\
$n4374_{(g-k)}$ & 59.8$\pm$26.2 & 30.2$\pm$26.2 & 2.920$\pm$0.061 & 2.659$\pm$0.249 & 3.632$\pm$0.512 & 0.565$\pm$0.041 & 0.408$\pm$0.118 & 0.363$\pm$0.165 & 0.810 & 0.745 & 0.548 \\ 
$n4374_{(z-k)}$ & 46.8$\pm$27.4 & 43.2$\pm$27.4 & 1.881$\pm$0.048 & 1.664$\pm$0.282 & 2.196$\pm$0.412 & 0.479$\pm$0.035 & 0.349$\pm$0.135 & 0.347$\pm$0.176 & 0.869 & 0.820 & 0.682 \\
\hline
$n4382_{(g-z)}$ & 31.8$\pm$19.0 & 25.2$\pm$19.0 & 1.019$\pm$0.023 & 0.896$\pm$0.095 & 1.228$\pm$0.165 & 0.177$\pm$0.015 & 0.096$\pm$0.044 & 0.101$\pm$0.061 & 0.493 & 0.710 & 0.452 \\
$n4382_{(g-k)}$ & 13.3$\pm$13.9 & 43.7$\pm$13.9 & 3.302$\pm$0.063 & 2.539$\pm$0.427 & 3.473$\pm$0.188 & 0.461$\pm$0.045 & 0.235$\pm$0.134 & 0.335$\pm$0.092 & 0.669 & 0.489 & 0.702 \\
$n4382_{(z-k)}$ & 14.7$\pm$16.9 & 42.3$\pm$16.9 & 2.284$\pm$0.063 & 1.498$\pm$0.533 & 2.483$\pm$0.261 & 0.480$\pm$0.047 & 0.212$\pm$0.150 & 0.362$\pm$0.129 & 0.599 & 0.034 & 0.699 \\
\hline
$n4406_{(g-z)}$ & 54.7$\pm$17.2 & 21.3$\pm$17.2 & 1.013$\pm$0.022 & 0.922$\pm$0.057 & 1.290$\pm$0.112 & 0.189$\pm$0.015 & 0.114$\pm$0.031 & 0.094$\pm$0.047 & 0.013 & 0.091 & 0.277 \\
$n4406_{(g-k)}$ & 39.4$\pm$25.4 & 36.6$\pm$25.4 & 3.236$\pm$0.059 & 2.890$\pm$0.334 & 3.658$\pm$0.394 & 0.494$\pm$0.042 & 0.307$\pm$0.136 & 0.333$\pm$0.151 & 0.970 & 0.816 & 0.490 \\
$n4406_{(z-k)}$ & 25.5$\pm$20.8 & 50.5$\pm$20.8 & 2.222$\pm$0.050 & 1.893$\pm$0.355 & 2.391$\pm$0.350 & 0.410$\pm$0.048 & 0.358$\pm$0.243 & 0.303$\pm$0.140 & 0.072 & 0.932 & 0.994 \\
\hline
$n4473_{(g-z)}$ & 31.8$\pm$9.7  & 23.2$\pm$9.7  & 1.024$\pm$0.027 & 0.910$\pm$0.037 & 1.198$\pm$0.087 & 0.167$\pm$0.017 & 0.066$\pm$0.023 & 0.117$\pm$0.036 & 0.002 & 0.601 & 0.293 \\
$n4473_{(g-k)}$ & 38.9$\pm$9.0  & 16.1$\pm$9.0  & 2.989$\pm$0.086 & 2.861$\pm$0.083 & 3.521$\pm$0.586 & 0.598$\pm$0.082 & 0.351$\pm$0.097 & 0.681$\pm$0.326 & 0.077 & 0.890 & 0.987 \\
$n4473_{(z-k)}$ & 40.9$\pm$10.5 & 14.1$\pm$10.5 & 1.964$\pm$0.071 & 1.884$\pm$0.197 & 2.465$\pm$0.690 & 0.508$\pm$0.078 & 0.317$\pm$0.102 & 0.636$\pm$0.372 & 0.022 & 0.926 & 0.998 \\
\hline
$n4486_{(g-z)}$ & 160.7$\pm$22.6&140.3$\pm$22.6 & 1.152$\pm$0.013 & 0.970$\pm$0.027 & 1.361$\pm$0.032 & 0.227$\pm$0.006 & 0.111$\pm$0.017 & 0.121$\pm$0.018 & $<$0.001 & 0.11 & $<$0.001\\
$n4486_{(g-k)}$ & 179.2$\pm$93.7&121.8$\pm$93.7 & 3.199$\pm$0.032 & 2.912$\pm$0.335 & 3.653$\pm$0.427 & 0.563$\pm$0.026 & 0.473$\pm$0.124 & 0.370$\pm$0.181 & 0.963 & 0.682 & 0.585 \\
$n4486_{(z-k)}$ & 147.7$\pm$38.1&153.3$\pm$38.1 & 2.047$\pm$0.026 & 1.996$\pm$0.129 & 2.087$\pm$0.039 & 0.423$\pm$0.025 & 0.467$\pm$0.147 & 0.324$\pm$0.151 & $<$0.001 & 0.92 & 1 \\
\hline
$n4526_{(g-z)}$ & 28.6$\pm$4.2  & 26.4$\pm$4.2  & 1.049$\pm$0.033 & 0.852$\pm$0.012 & 1.263$\pm$0.027 & 0.223$\pm$0.013 & 0.045$\pm$0.008 & 0.120$\pm$0.025 & $<$0.001 & 0.05 & 0.001 \\
$n4526_{(g-k)}$ & 29.1$\pm$15.9 & 25.9$\pm$15.9 & 3.244$\pm$0.082 & 3.001$\pm$0.352 & 3.671$\pm$0.540 & 0.596$\pm$0.055 & 0.351$\pm$0.162 & 0.510$\pm$0.251 & 0.575 & 0.893 & 0.828 \\
$n4526_{(z-k)}$ & 27.7$\pm$13.6 & 27.3$\pm$13.6 & 2.195$\pm$0.068 & 2.037$\pm$0.260 & 2.467$\pm$0.427 & 0.517$\pm$0.055 & 0.262$\pm$0.113 & 0.542$\pm$0.209 & 0.287 & 0.915 & 0.908 \\
\hline
$n4552_{(g-z)}$ & 61.2$\pm$21.3 & 45.8$\pm$21.3 & 1.139$\pm$0.019 & 0.992$\pm$0.060 & 1.348$\pm$0.088 & 0.212$\pm$0.011 & 0.115$\pm$0.035 & 0.125$\pm$0.043 & 0.011 & 0.247 & $<$0.001\\ 
$n4552_{(g-k)}$ & 56.9$\pm$32.1 & 50.1$\pm$32.1 & 3.106$\pm$0.055 & 2.853$\pm$0.397 & 3.420$\pm$0.415 & 0.510$\pm$0.038 & 0.382$\pm$0.179 & 0.377$\pm$0.167 & 0.535 & 0.884 & 0.846 \\
$n4552_{(z-k)}$ & 58.8$\pm$20.8 & 48.2$\pm$20.8 & 1.967$\pm$0.047 & 1.914$\pm$0.135 & 2.090$\pm$0.327 & 0.412$\pm$0.040 & 0.255$\pm$0.145 & 0.486$\pm$0.199 & $<$0.001 & 0.964 & 0.999 \\
\hline
$n4570_{(g-z)}$ & -              &        -     &        -        &         -       &          -      &          -      &          -      &        -        &        - &       &       \\
$n4570_{(g-k)}$ & -              &        -     &        -        &         -       &          -      &          -      &          -      &        -        &        - &       &       \\
$n4570_{(z-k)}$ & -              &        -     &        -        &         -       &          -      &          -      &          -      &        -        &        - &       &       \\
\hline
$n4621_{(g-z)}$ & 32.5$\pm$5.5  & 43.5$\pm$5.5  & 1.162$\pm$0.027 & 0.952$\pm$0.022 & 1.318$\pm$0.023 & 0.20$\pm$0.011  & 0.084$\pm$0.013 & 0.107$\pm$0.018 & $<$0.001 & 0.085 & 0.001 \\
$n4621_{(g-k)}$ & 24.4$\pm$18.2 & 51.6$\pm$18.2 & 3.215$\pm$0.065 & 2.592$\pm$0.482 & 3.383$\pm$0.108 & 0.493$\pm$0.045 & 0.347$\pm$0.171 & 0.333$\pm$0.105 & 0.429 & 0.839 & 0.857 \\
$n4621_{(z-k)}$ & 27.8$\pm$18.9 & 48.2$\pm$18.9 & 2.054$\pm$0.047 & 1.867$\pm$0.288 & 2.149$\pm$0.196 & 0.368$\pm$0.047 & 0.485$\pm$0.203 & 0.244$\pm$0.124 & 0.01 & 0.92 & 0.999 \\
\hline
$n4649_{(g-z)}$ & 76.4$\pm$18.8 & 84.6$\pm$8.8  & 1.192$\pm$0.015 & 0.969$\pm$0.021 & 1.394$\pm$0.027 & 0.243$\pm$0.009 & 0.105$\pm$0.017 & 0.129$\pm$0.016 & $<$0.001 & 0.089 &$<$0.001 \\
$n4649_{(g-k)}$ & 99.5$\pm$26.7 & 61.5$\pm$26.7 & 3.195$\pm$0.043 & 2.857$\pm$0.185 & 3.723$\pm$0.141 & 0.582$\pm$0.025 & 0.457$\pm$0.074 & 0.279$\pm$0.085 & 0.037 & 0.474 & 0.021 \\ 
$n4649_{(z-k)}$ & 76.4$\pm$41.0 & 84.6$\pm$41.0 & 2.003$\pm$0.033 & 1.760$\pm$0.307 & 2.144$\pm$0.110 & 0.420$\pm$0.021 & 0.402$\pm$0.133 & 0.270$\pm$0.100 & 0.341 & 0.817 & 0.872 \\
\hline
$n4660_{(g-z)}$ & 38.9$\pm$7.9  & 10.1$\pm$7.9  & 0.946$\pm$0.020 & 0.895$\pm$0.031 & 1.183$\pm$0.094 & 0.136$\pm$0.016 & 0.087$\pm$0.017 & 0.076$\pm$0.040 & 0.085 & 0.549 & 0.824 \\
$n4660_{(g-k)}$ & 38.1$\pm$12.4 & 10.9$\pm$12.4 & 2.979$\pm$0.055 & 2.860$\pm$0.141 & 3.678$\pm$0.417 & 0.385$\pm$0.043 & 0.277$\pm$0.075 & 0.209$\pm$0.135 & 0.479 & 0.195 & 0.904 \\
$n4660_{(z-k)}$ & 29.9$\pm$15.5 & 19.1$\pm$15.5 & 2.033$\pm$0.050  & 1.87$\pm$0.307 &  2.324$\pm$0.398 & 0.339$\pm$ 0.057 & 0.249$\pm$0.169 & 0.393$\pm$0.263 & 0.012 & 0.923& 1 \\
\hline
\hline

\end{tabular}
\caption{GMM outputs for the colour distributions of the galaxies: (1) galaxy and colour distributions, (2) number of clusters assigned to the blue peak ($N_{b}$),  (3) number of clusters assigned to the red peak ($N_{r}$), (4) the mean of the unimodal distribution ($\mu$), (5) the mean of the blue peak ($\mu_{b}$),  (6) the mean of the red peak ($\mu_{r}$),  (7) the width of the unimodal distribution ($\sigma$), (8) the width of the blue peak ($\sigma_{b}$), (9) the width of the red peak ($\sigma_{b}$) and (10), (11), (12) the probabilities of the different statistics for rejecting a unimodal distribution P($\chi^{2}$), P(DD), P(kurt).}
\label{gmm}
\end {scriptsize}
\end{table}
\end{landscape}

\begin{figure}
\includegraphics[width=8cm]{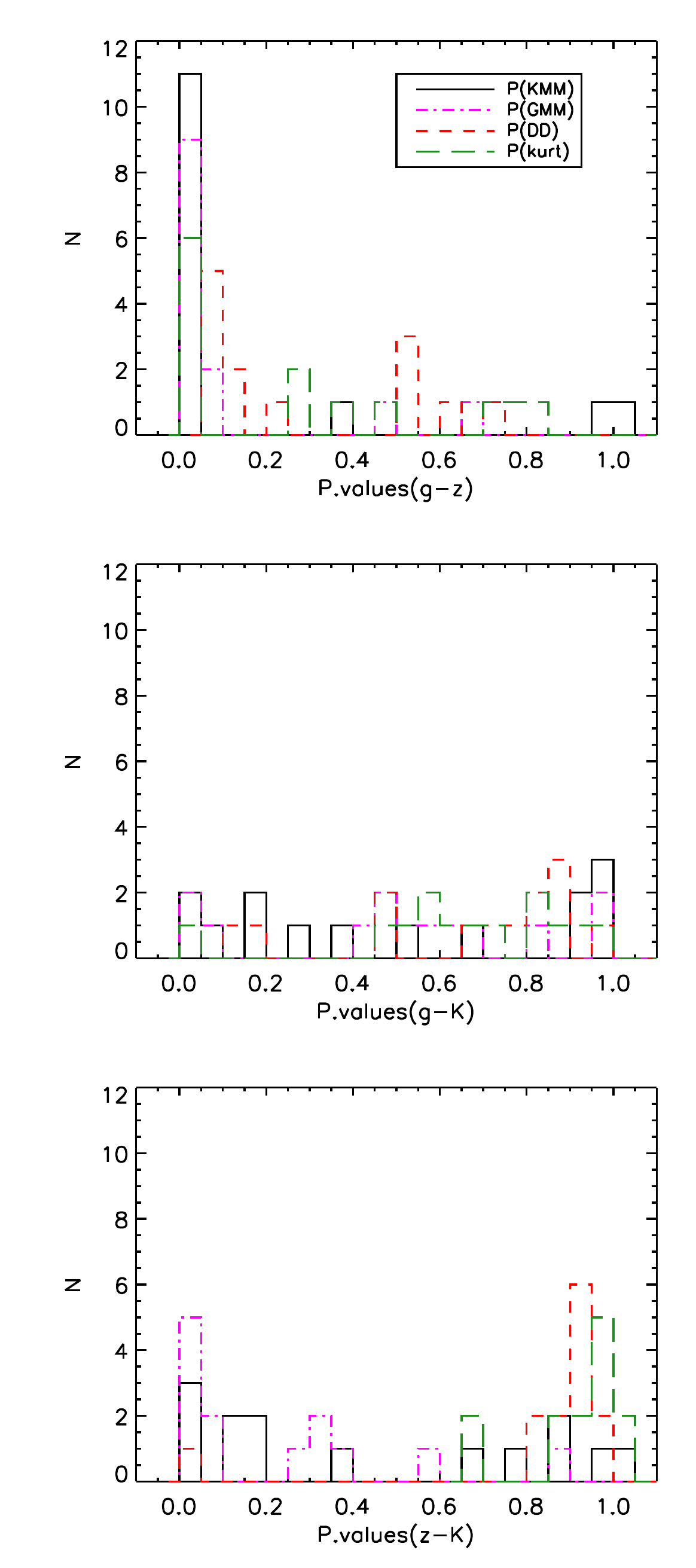}
\caption{$(g-z)$ (\textit{top panel}), $(g-K)$(\textit{middle panel}) and $(z-K)$ (\textit{bottom panel}) probability values from KMM and GMM.}
\label{pvaluesgal}
\end{figure}

In Fig. \ref{pvaluesgal}, histograms of the different probabilities returned by KMM and GMM are shown.
The $(g-z)$ probabilities of these tests are concentrated towards 0 rather than 1, indicating the probable rejection of the unimodal distribution.
In fact, $\sim 80\%$ of the $(g-z)$ distributions of P(KMM) and $\sim 70\%$ of P(GMM) are consistent with $\le$ 0.05.  
The P(DD) and P(kurt) are $\le$ 0.1 for $\sim 40\%$ and $\sim 46\%$ of the systems, respectively.
Therefore a bimodal distribution is favoured for approximately half of the cases when the optical colour is used.
In contrast, the $(g-K)$ probability values are spread between 0 and 1.
Only $\sim 15\%$ of the cases have P(KMM) and P(GMM) $\le$ 0.05.
The P(kurt) is $\le$ 0.1 for only $\sim 8\%$ of the systems, while none have P(DD) $\le$ 0.1.
Like $(g-K)$, the $(z-K)$ probability values are also spread between 0 and 1. 
Compared to the former colour, a slightly larger number of systems have P(KMM) and P(GMM) $\le$ 0.05 ($\sim21\%$ and $\sim38\%$, respectively).
The P(DD) is $\le$ 0.1 for only $\sim 8\%$ of the systems, while none have P(kurt) $\le$ 0.1.
Note however, that generally for $(z-K)$ and for a few cases of $(g-K)$ these tests assign very few objects for one peak compared to the other, while in $(g-z)$ the number assigned to the different peaks is much more equal.
In summary, the higher probability values assigned by KMM and GMM for $(g-K)$ and $(z-K)$ attest that bimodality does get less evident in $(g-K)$ and $(z-K)$ compared to $(g-z)$.

\section{The cluster rich galaxies and simulations of the colour distributions}
The $(z-K)$ and $(g-K)$ colours, where bimodality appears generally less pronounced, are also the colours with the higher photometric uncertainties (\citealt{paper1}). 
In the present section we investigate whether photometric errors might be the reason bimodality is less apparent in such colours. 

\begin{figure*}[h]
\centering
\includegraphics[width=60mm]{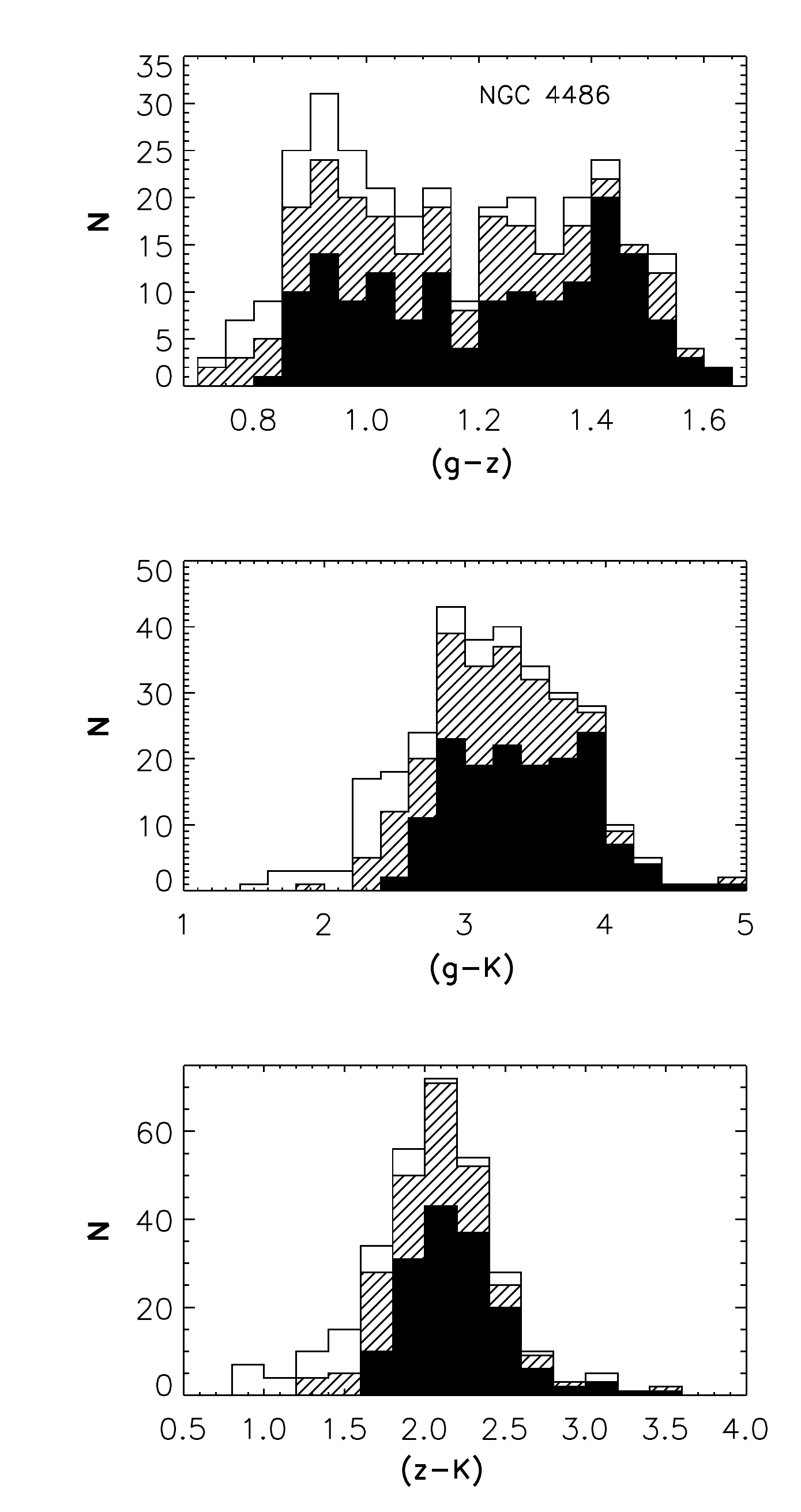}
\includegraphics[width=60mm]{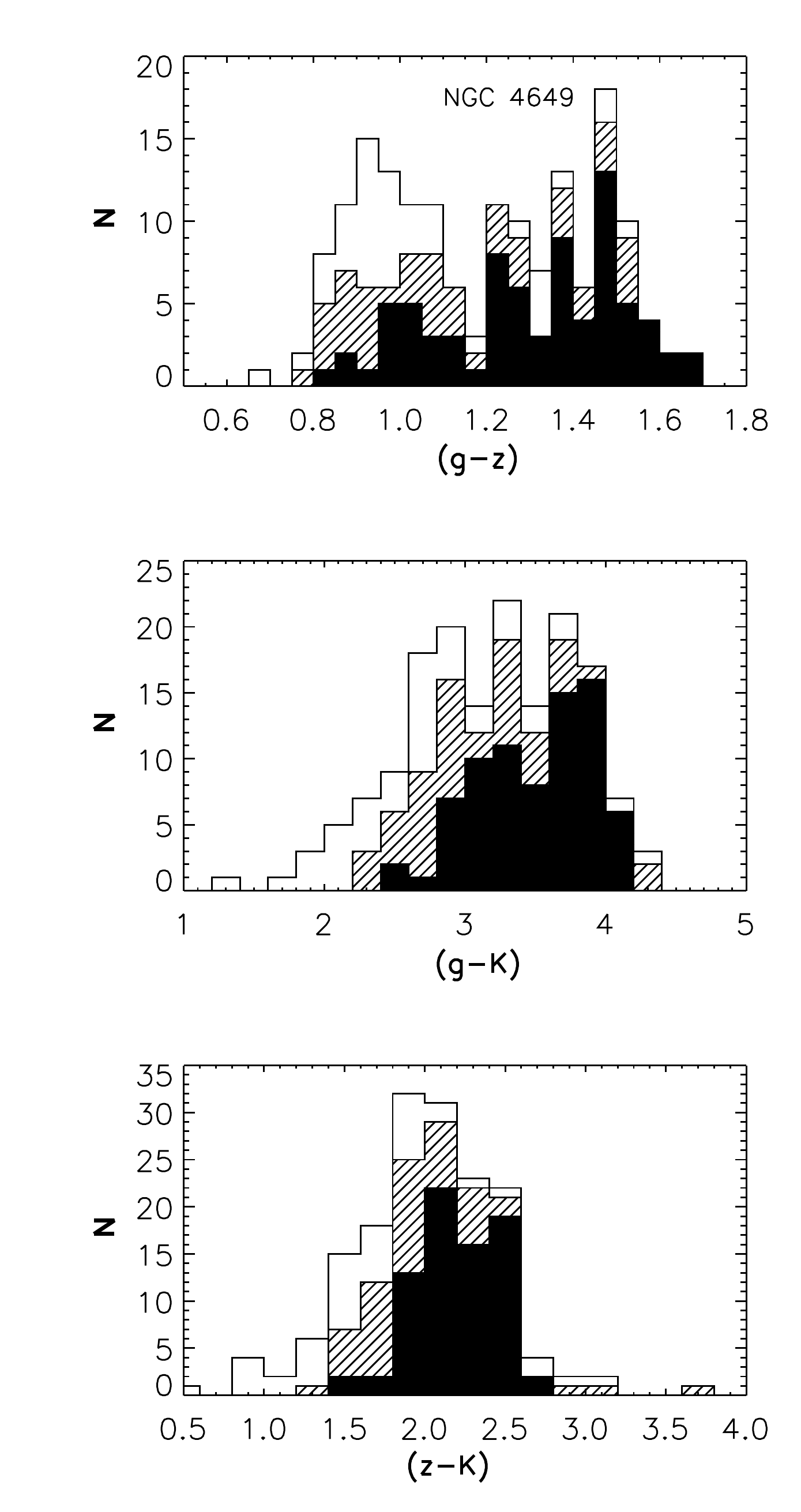}
\includegraphics[width=60mm]{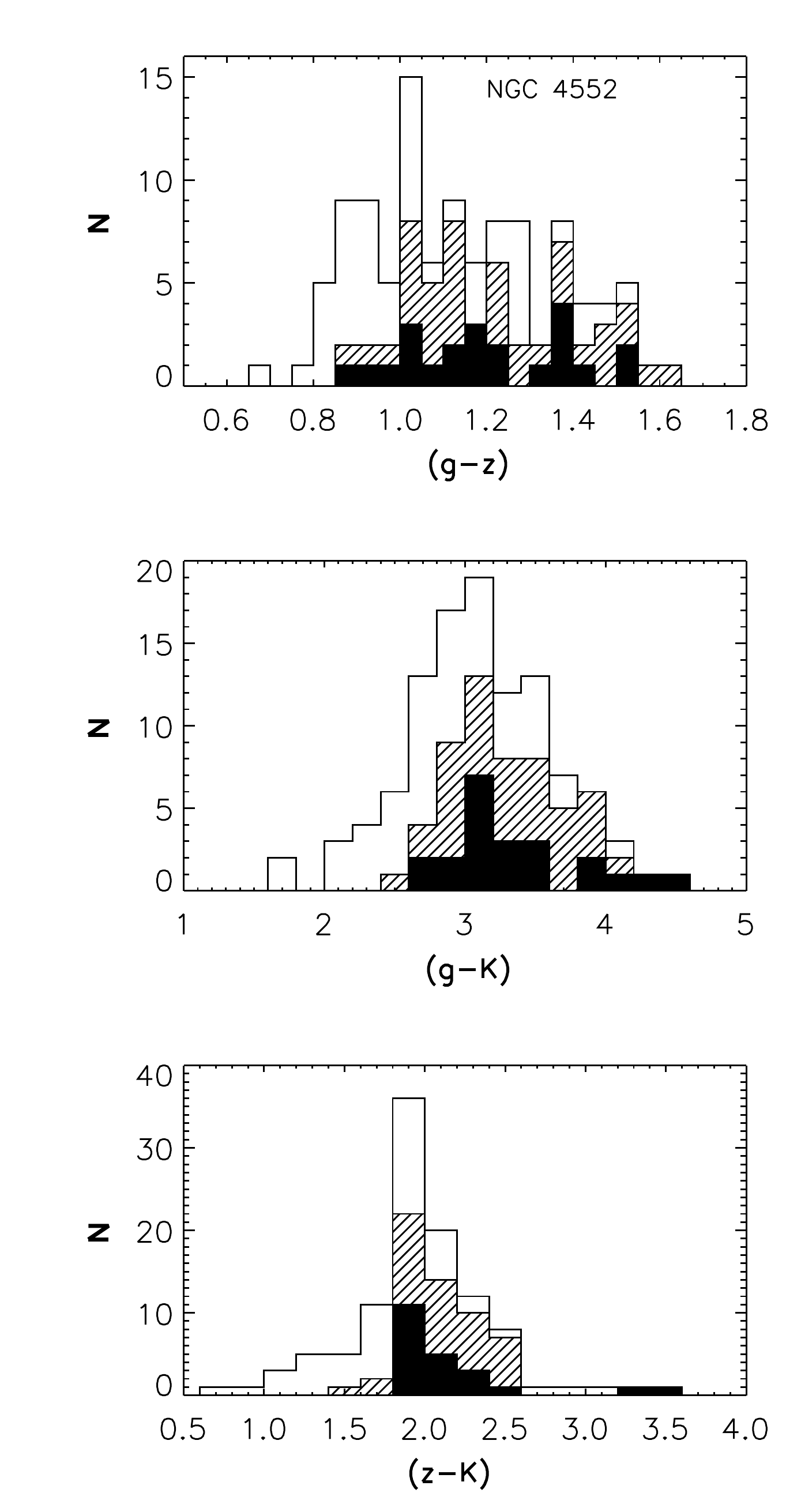}
\caption{NGC\,4486, NGC\,4649 and NGC\,4552 colour distributions in $(g-z)$, $(g-k)$ and $(z-k)$. The open histogram are the colour distributions for all clusters that make the final sample, according to the criteria outlined is section 2. The hashed histograms show the same distributions when only the clusters with $K_{err}\le 0.1$ are left in the sample and the filled histogram when only the clusters with $K_{err}\le 0.05$ are present.}
\label{clustrichcoldisits}
\end{figure*}

First we take the three richest GC systems (NGC\,4486, NGC\,4649 and NGC\,4552) to see how the distributions look when samples with restricted photometric errors are considered.  
The colour distributions for these cluster rich galaxies are shown in Fig. \ref{clustrichcoldisits} for different ranges of photometric uncertainties: $K_{err}\le 0.5$ (whole sample), $K_{err} \le 0.1$ and $K_{err}\le 0.05$. The outputs of KMM and GMM are shown in Tabs. \ref{clustrichtabkmm} and \ref{clustrichtabgmm} respectively. When applying these restrictions, a greater fraction of blue objects drops off the sample, as compared to red objects. This is due to the colour dependant scatter, as discussed in \cite{paper2}.
For the $(g-z)$ distribution, all probability values are significantly low, indicating bimodality both for samples with $K_{err} \le 0.1$ and $K_{err}\le 0.05$. 
For the other colours the situation changes and distinguishing a unimodal distribution from a bimodal one through KMM and GMM becomes very difficult.
It can happen that a probability value goes from $\sim$ 1 to $\sim$ 0 with the restricted sample, e.g. P(GMM) of $(g-K)$ for NGC\,4486. Also, it happens that most objects are assigned to one of the peaks in the $(g-K)$ and $(z-K)$ KMM estimates of the $K_{err} \le 0.1$ and $K_{err}\le0.5$ samples (e.g. for NGC\,4649). 
This exercise does not lead to a clear conclusion regarding the presence of bimodality
in the optical/NIR colour distributions, mainly due to the low number statistics when running the tests in the samples restricted in photometric uncertainties. 
As generally assumed, if the $(g-z)$ distribution is clearly bimodal as it is for NGC\,4486 one would expect to see bimodal $(g-K)$ and $(z-K)$ distributions in the $K_{err}\le 0.05$ sample.
Instead, the optical/NIR distributions look much less bimodal.
However, note that bimodality is somewhat significant in $(g-K)$ for NGC\,4486 for the $K_{err}\le 0.05$ sample, with P(KMM)$=0.035$ and P(GMM)$<0.001$.

\subsection{Simulations including photometric scatter} 
Here, we test whether bimodality could be blurred in $(g-K)$ and $(z-K)$ due to the high photometric uncertainties in $K$.
As shown in \cite{blakeslee10} the most subtle details in the colour metallicity relation matter for the final transformed distribution. 
Therefore, to avoid unnecessary complications we do not use SSP models to transform the colours but assume linear transformations. A bimodal distribution (in this case, $(g-z)$) should maintain its form if linearly mapped to other colour spaces. 
We find the relations between $(g-z)$ and $(g-K)$, and between $(g-z)$ and $(z-K)$ by fitting the best straight line by a procedure that gives more weight to the objects with less observational errors in both the abscissa and the ordinate using the GC systems of NGC\,4486 and NGC\,4649, as in \cite{paper2}. The relations are given by:

\begin{equation}
\centering
(g-z)=0.465*(g-K)-0.349
\end{equation} 

\begin{equation}
\centering
(g-z)=0.746*(z-K)-0.387
\end{equation} 
 
\begin{figure}
\centering
\includegraphics[width=8cm]{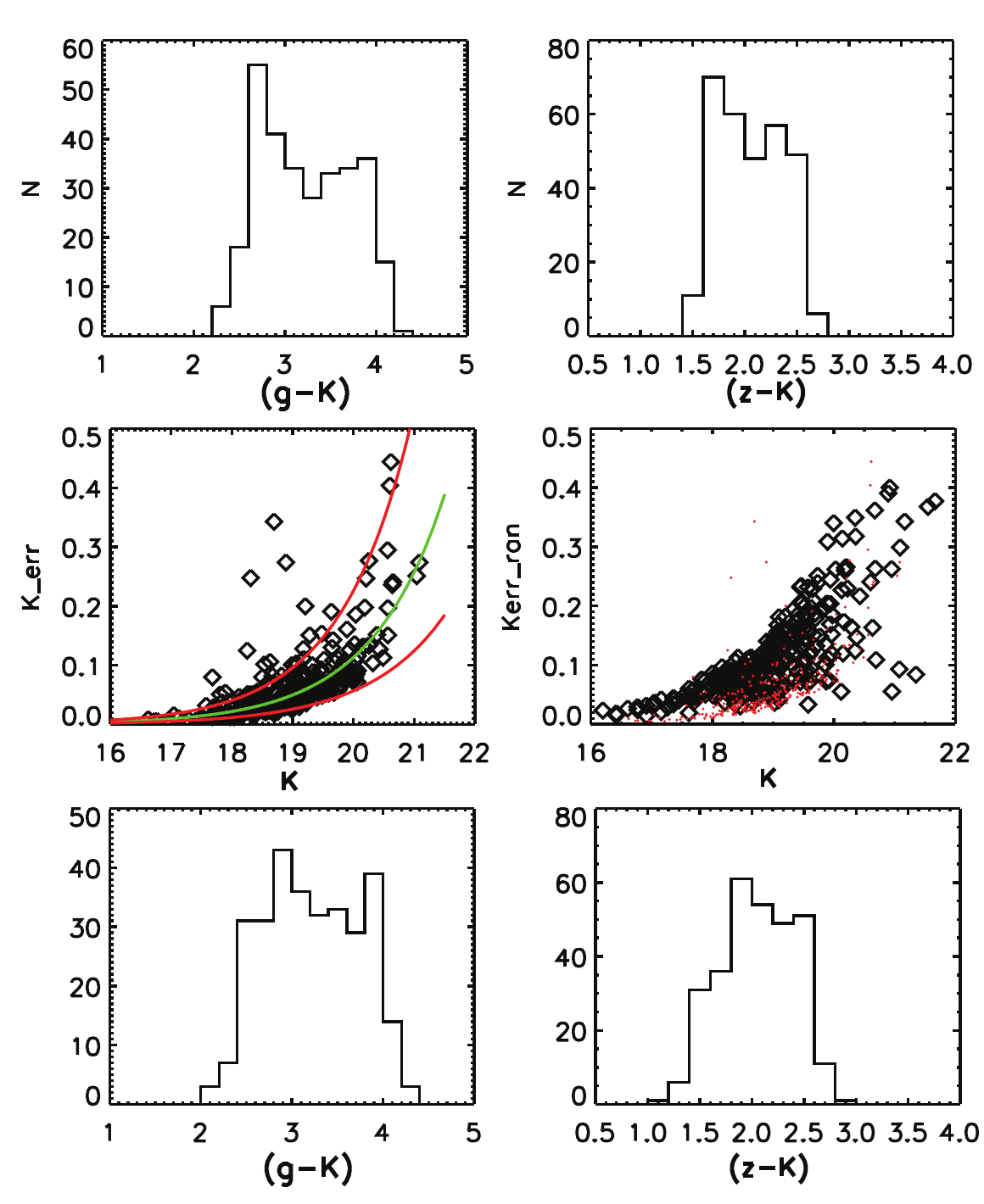}
\caption{Illustration for NGC\,4486 on how the simulations with realistic photometric errors were performed. \textit{Top panels}: the $(g-K)$ and $(z-K)$ colour distributions transformed linearly from $(g-z)$ using relations (1) and (2) respectively. \textit{Middle panels}: the data and the modelled photometric scatter ($K_{err}$) as a function of K and one outcome of the randomly sampled photometric scatter ($K_{err_{ran}}$) as a function of K. \textit{Bottom panels}: the $(g-K)$ and $(z-K)$ simulations with $K_{err_{ran}}$ added.}
\label{simulex}
\end{figure}

We linearly transform the $(g-z)$ distribution to $(g-K)$ and $(z-K)$
using the inverted relations (1) and (2). In the top
panels of Fig. \ref{simulex} the results of this linear transformation
are shown for NGC\,4486.  Recalling the result of \cite{paper1} that
the photometric scatter in $K$, as measured by PHOT, is underestimated
by a factor of 2, we add randomly a realistic $K$ dependent scatter to
the $(g-K)$ and $(z-K)$ transformed distributions. The contributions
of the $g$ and $z$ errors to these colours are negligible if compared
to the $K$ errors.  In order to add the error associated to each
$K$-band magnitude to the colours we first determine an analytic
relation between $K$ \textit{vs.} $K_{err}$.  The middle left panel of
Fig.\,\ref{simulex} shows the observed $K$ and $K_{err}$, the best fit
relation as well as its one sigma deviation.  The analytic relation is
of the form $K_{err}=N\times10^{a*K}$, where $N$ and $a$ are constants
that are different for different galaxies.  
A random number drawn from a gaussian with dispersion corresponding to the measured one sigma deviation of the $K$ \textit{vs.} $K_{err}$ relation is calculated. As in \cite{paper1} it was shown that the photometric errors are underestimated by a factor of 2, we multiply this random number by two and add it to the analytic $K_{err}$. Following this procedure we associate a $K$-band error to each GC according to its magnitude. 

One outcome of the randomly sampled realistic scatter is shown in the middle right panel of Fig. \ref{simulex} as a scatter plot of large black symbols. For comparison, the parent distribution of errors is also shown as a scatter plot of small red dots.
In the bottom panels the resulting $(g-K)$ and $(z-K)$ distributions after the scatter is added are shown. 
A realistic scatter is sampled randomly and added to the linearly transformed distribution 100 times to generate 100 unique distributions.
In order to have a sufficient number of clusters for a robust analysis, we do this only for the most GC-rich galaxies that show obvious bimodality in $(g-z)$ and a nearly equal number of clusters in the blue and red peaks: NGC\,4486, NGC\,4649, NGC\,4552 and NGC\,4621.

\begin{figure}
\centering
\includegraphics[width=8cm]{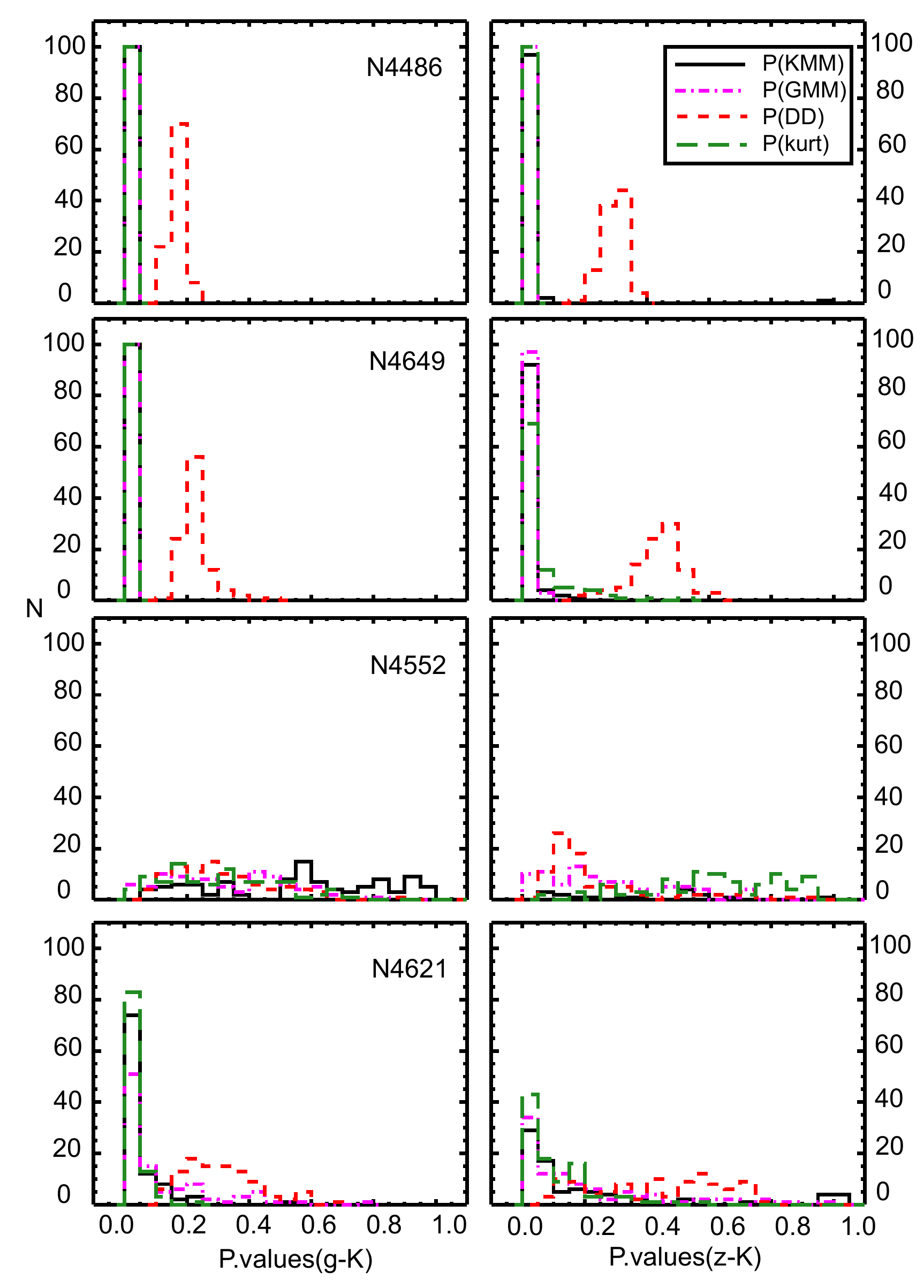}
\caption{Histograms of the probability values returned by KMM and GMM on the output of the simulations for the cluster rich galaxies with obvious $(g-z)$ bimodality and nearly equal number of clusters assigned for the blue and red peaks; for $(g-K)$ (\textit{left panels}) and $(z-K)$ (\textit{right panels}).}
\label{outpval_simuls}
\end{figure}

KMM and GMM ere run on the outputs of these simulations. In Fig. \ref{outpval_simuls}, histograms of the probability values returned by these statistical tests are shown.
For NGC\,4486 and NGC\,4649 a linearly transformed bimodal distribution in $(g-z)$ remains bimodal both in $(g-K)$ and $(z-K)$ even when realistic photometric scatter is added. This is attested by the statistically significant values of P(KMM), P(GMM) and P(kurt) ($\le$ 0.05) for the great majority of the cases. Nonetheless, the separation between the peaks is not as significant, P(DD) $>$ 0.05.
However, for NGC\,4552 and NGC\,4621 the high probability values show that photometric scatter could blur bimodality in $(g-K)$ and $(z-K)$ in many of the outcomes. Moreover, for NGC\,4552, KMM does not converge for many cases, probably because of the low number of clusters.  
Therefore, the only galaxies for which we can robustly analyse the optical/NIR colour distributions are NGC\,4486 and NGC\,4649. 

This analysis shows that if unimodality is preferred over bimodality in $(g-K)$ and $(z-K)$ it is not an effect of photometric errors in NGC\,4486 and NGC\,4649.
KMM and GMM tests show that their $(g-z)$ distributions are genuinely bimodal. The $(g-K)$ distribution is unimodal for NGC\,4486 while bimodal for NGC\,4649.
For the reddest of the colours, $(z-K)$, bimodality is not significant for any of the galaxies.

{\small
\begin{table*}
\centering
\begin{tabular}{c c ccc  ccc}

\hline\hline
Galaxy & $\mu_{blue}$ & $\mu_{red}$  &  $N_{blue}$  & $N_{red}$ & $\sigma$& $P(KMM)$ \\
\hline
$n4486_{(g-z)}$($K_{err} \le 0.1$)   & 0.99 & 1.37     &  131      &  122     &       0.11    &  0.000     \\
$n4486_{(g-z)}$($K_{err} \le 0.05$) &  1.01   & 1.39      &    69     &  85     &    0.10        &    0.000         \\
$n4486_{(g-k)}$($K_{err} \le 0.1$)   &3.03 & 3.69     &  155      &  98     &       0.37    &0.324    \\
$n4486_{(g-k)}$($K_{err} \le 0.05$) & 3.08   & 3.80      &    82    &  72     &    0.30        &    0.035       \\
$n4486_{(z-k)}$($K_{err} \le 0.1$)   &2.09 & 3.10    &  245      &  8     &       0.21    &0.324    \\
$n4486_{(z-k)}$($K_{err} \le 0.05$) &2.16& 3.06    &    147    &  7     &    0.25        &    0.000        \\
\hline
$n4649_{(g-z)}$($K_{err} \le 0.1$)  &  1.00   & 1.41   &  49       &   74    &    0.13          &  0.000        \\
$n4649_{(g-z)}$($K_{err} \le 0.05$)&  1.04   &  1.42     &   21      &  56     &     0.12         &    0.002      \\
$n4649_{(g-k)}$($K_{err} \le 0.1$)  &    3.34 &   5.20    &   121      &  2     &      0.49        &      0.030    \\
$n4649_{(g-k)}$($K_{err} \le 0.05$)& 3.47    &   3.61    &     71      &     1 &0.46        &   0.999       \\
$n4649_{(z-k)}$($K_{err} \le 0.1$)  &   2.10  &    4.15   &   121      &  2     &    0.32         &     0.000    \\
$n4649_{(z-k)}$($K_{err} \le 0.05$)&   2.19  &    2.25   &    76    &   1    &      0.37        &      0.999     \\
\hline
$n4552_{(g-z)}$($K_{err} \le 0.1$)  &   1.18  &   1.23    &   19      &  39     &     0.19       &     1     \\
$n4552_{(g-z)}$($K_{err} \le 0.05$)& 1.07    &  1.38    &   14     &   8    &       0.10       &       0.270   \\
$n4552_{(g-k)}$($K_{err} \le 0.1$)  &   3.13  &  3.84     &    44     &  14     &    0.30          &  0.124        \\
$n4552_{(g-k)}$($K_{err} \le 0.05$)&  3.150  &   4.12    &  17       &  5     &        0.26      & 0.039         \\
$n4552_{(z-k)}$($K_{err} \le 0.1$)  &   2.07  &  3.40     &   56     &  2    &    0.23          & 0.000        \\
$n4552_{(z-k)}$($K_{err} \le 0.05$)&  2.04  &  3.40   &  20       &  2     &        0.18    &0.000         \\
\hline
\end{tabular}
\caption{Results of KMM for NGC\,4486, NGC\,4649 and NGC\,4552 for the colour distributions with GCs $K_{err} \le 0.1$ and $K_{err} \le 0.05$: (1) galaxy, (2) number of clusters assigned to the blue peak ($N_{b}$), (3) number of clusters assigned to the red peak ($N_{r}$), (4) the mean of the blue peak ($\mu_{b}$), (5) the mean of the red peak ($\mu_{r}$), (6) the common width of the peaks ($\sigma$) and (7) the probability for rejecting a unimodal distribution (P(KMM)).}
\label{clustrichtabkmm}
\end{table*}
}

\begin{landscape}

\begin{table}
\begin{scriptsize}
\centering
\begin{tabular}{cccccccccccccccccc}
\hline
             (1)  & (2) & (3)   & (4)     &      (5)   &  (6)  &  (7) & (8) & (9) & (10) & (11) & (12)\\
\hline
      Galaxy & $N_{b}$ & $N_{r}$ & $\mu$ & $\mu_{b}$  &$\mu_{r}$ & $\sigma$ & $\sigma_{b}$ &	$\sigma_{b}$ & P($\chi^{2}$) & P(DD) & P(kurt)  \\
$n4486_{(g-z)}$($K_{err} \le 0.1$) & 123.4$\pm$25.2 & 129.6$\pm$25.2 & 1.174$\pm$0.015 & 0.981$\pm$0.036 & 1.359$\pm$0.040 & 0.221$\pm$0.006 & 0.104$\pm$0.023 & 0.124$\pm$0.023 & $<$0.001 & 0.117 & $<$ 0.001 \\
$n4486_{(g-z)}$($K_{err} \le 0.05$) &67.3$\pm$13.0  & 86.7$\pm$13.0  & 1.221$\pm$0.018 & 1.005$\pm$0.034 & 1.389$\pm$0.032 & 0.217$\pm$0.007 & 0.098$\pm$0.021 & 0.108$\pm$0.023 & $<$0.001 & 0.090 & $<$0.001 \\
$n4486_{(g-k)}$($K_{err} \le 0.1$) & 118.6$\pm$80.0 & 134.4$\pm$80.0 & 3.306$\pm$0.032 & 2.962$\pm$0.194 & 3.822$\pm$0.576 & 0.493$\pm$0.022 & 0.281$\pm$0.117 & 0.352$\pm$0.149 & 0.125 & 0.638 & 0.565 \\
$n4486_{(g-k)}$($K_{err} \le 0.05$)& 42.6$\pm$22.2  & 111.4$\pm$22.2 & 3.426$\pm$0.039 & 2.899$\pm$0.106 & 3.642$\pm$0.194 & 0.473$\pm$0.023 & 0.151$\pm$0.057 & 0.388$\pm$0.071 & $<$0.001 & 0.360 & 0.233\\  
$n4486_{(z-k)}$($K_{err} \le 0.1$) & 227.3$\pm$37.6 & 25.7$\pm$37.6  & 2.132$\pm$0.022 & 2.089$\pm$0.026 & 2.951$\pm$0.407 & 0.346$\pm$0.022 & 0.276$\pm$0.038 & 0.292$\pm$0.139 & $<$0.001 & 0.104 & 1.000 \\
$n4486_{(z-k)}$($K_{err} \le 0.05$)& 138.1$\pm$22.6 & 15.9$\pm$22.6  & 2.205$\pm$0.025 & 2.141$\pm$0.052 & 2.995$\pm$0.268 & 0.324$\pm$0.023 & 0.245$\pm$0.031 & 0.226$\pm$0.096 & $<$0.001 & 0.082 & 0.999\\
\hline
$n4649_{(g-z)}$($K_{err} \le 0.1$)  & 60.8$\pm$17.8  & 62.2$\pm$17.8  & 1.236$\pm$0.023 & 1.029$\pm$0.067 & 1.433$\pm$0.047 & 0.236$\pm$0.009 & 0.135$\pm$0.037 & 0.105$\pm$0.033 & $<$0.001 & 0.156 & $<$0.001\\
$n4649_{(g-z)}$($K_{err} \le 0.05$)&22.1$\pm$9.4   & 54.9$\pm$9.4   & 1.307$\pm$0.026 & 1.025$\pm$0.069 & 1.416$\pm$0.039 & 0.210$\pm$0.013 & 0.092$\pm$0.038 & 0.124$\pm$0.026 & 0.009 & 0.122 & 0.023 \\
$n4649_{(g-k)}$($K_{err} \le 0.1$)  &61.5$\pm$0.0   & 61.5$\pm$0.0   & 3.363$\pm$0.052 & 3.363$\pm$0.052 & 3.363$\pm$0.052 & 0.540$\pm$0.047 & 0.540$\pm$0.047 & 0.540$\pm$0.047 & 0.827 & 1.000 & 0.970\\
$n4649_{(g-k)}$($K_{err} \le 0.05$)&44.6$\pm$23.5  & 32.4$\pm$23.5  & 3.528$\pm$0.055 & 3.319$\pm$0.280 & 4.162$\pm$0.756 & 0.451$\pm$0.052 & 0.321$\pm$0.137 & 0.240$\pm$0.184 & 0.014 & 0.001 & 0.999 \\
$n4649_{(z-k)}$($K_{err} \le 0.1$)  & 116.6$\pm$8.8  & 6.4$\pm$8.8    & 2.127$\pm$0.040 & 2.079$\pm$0.040 & 3.642$\pm$0.770 & 0.401$\pm$0.077 & 0.308$\pm$0.023 & 0.284$\pm$0.270 & $<$0.001 & 0.150 & 1.000 \\
$n4649_{(z-k)}$($K_{err} \le 0.05$)&64.4$\pm$15.7  & 12.6$\pm$15.7  & 2.221$\pm$0.043 & 2.132$\pm$0.120 & 3.656$\pm$1.123 & 0.339$\pm$0.098 & 0.224$\pm$0.042 & 0.073$\pm$0.042 & $<$0.001 & $<$0.001 & 1.000 \\
\hline
$n4552_{(g-z)}$($K_{err} \le 0.1$)  &36.4$\pm$6.4   & 21.6$\pm$6.4   & 1.214$\pm$0.027 & 1.091$\pm$0.030 & 1.426$\pm$0.048 & 0.190$\pm$0.013 & 0.103$\pm$0.022 & 0.091$\pm$0.026 & 0.017 & 0.171 & 0.010 \\
$n4552_{(g-z)}$($K_{err} \le 0.05$)&29.0$\pm$0.0   & 29.0$\pm$0.0   & 3.333$\pm$0.060 & 3.333$\pm$0.060 & 3.333$\pm$0.060 & 0.434$\pm$0.037 & 0.434$\pm$0.037 & 0.434$\pm$0.037 & 0.884 & 1.000 & 0.596  \\
$n4552_{(g-k)}$($K_{err} \le 0.1$)  &  &                &                 &                 &                 &                 &                 &                 &        &         &   \\
$n4552_{(g-k)}$($K_{err} \le 0.05$)&   &                &                 &                 &                 &                 &                 &                 &        &         &   \\
$n4552_{(z-k)}$($K_{err} \le 0.1$)  & 12.4$\pm$4.4   & 9.6$\pm$4.4    & 3.382$\pm$0.108 & 3.065$\pm$0.096 & 3.877$\pm$0.309  & 0.472$\pm$0.077 & 0.167$\pm$0.062 & 0.351$\pm$0.139 & 0.111 & 0.741 & 0.774 \\
$n4552_{(z-k)}$($K_{err} \le 0.05$)&   &                &                 &                 &                 &                 &                 &                 &        &         &   \\

\hline
\end{tabular}
\caption{Results of GMM for NGC\,4486, NGC\,4649 and NGC\,4552 for the colour distributions with GCs $K_{err} \le 0.1$ and $K_{err} \le 0.05$: columns as in Tab. \ref{gmm}.}
\label{clustrichtabgmm}
\end{scriptsize}
\end{table}

\end{landscape}

\subsection{Simulations accounting for scatter in the colour-colour diagrams}
The analysis performed above is intended to reproduce
  the observed K--K$_{err}$ parameter space, and showed that the
  photometric uncertainty does not play a major role in ``smoothing''
  the observed $(g-K)$ and $(z-K)$ colour distributions. However, it does
  not take into account the scatter in the $(g-z)$ \textit{vs.}
  $(g-K)$ and $(g-z)$ \textit{vs.} $(z-K)$ diagrams, that might be an
  intrinsic scatter, at least in part.

In order to take into account the effect of the scatter in the
colour-colour diagrams we performed two more sets of simulations for
NGC\,4486 and NGC\,4649. For the second set of simulations, as before,
we linearly transformed the $(g-z)$ colour to $(g-K)$ and $(z-K)$
using Eqs (1) and (2). Following this, we calculated the residual for
each GC from the transformation relations (1) and (2).  For each
galaxy we then simulated 100 colour distributions (both for $(g-K)$
and $(z-K)$) by adding errors randomly pulled from the array of
residuals with replacement to the transformed $(g-K)$ and $(z-K)$
disitributions.

In the second paper of this series (see Fig. 14,
  \citealt{paper2}) it is shown that the scatter estimated through a
  broken fit in the colour-colour diagrams (rather than a linear fit)
  is almost entirely consistent with the photometric scatter. However,
  there we also note that there is still extra scatter found at bright
  magnitudes that  may be attributed to age spreads or other factors.  Following this, we performed
  a third set of simulations that consisted in using broken fit
  relations for transforming the $(g-z)$ colour to $(g-K)$ and
  $(z-K)$.  These are given by Eqs. (3), (4), (5) and (6). Eqs. (3)
  and (5) are valid for $(g-z)<1.187$ and Eqs. (4) and (6) for
  $(g-z)>1.187$.

\begin{equation}
\centering
(g-z)_b=0.260*(g-K)_b+0.232
\end{equation} 

\begin{equation}
\centering
(g-z)_r=0.340*(g-K)_r+0.140
\end{equation} 

\begin{equation}
\centering
(g-z)_b=0.292*(z-K)_b+0.431
\end{equation} 

\begin{equation}
\centering
(g-z)_r=0.437*(z-K)_r+0.391
\end{equation} 

\begin{figure}
\centering
\includegraphics[width=6cm, angle=90]{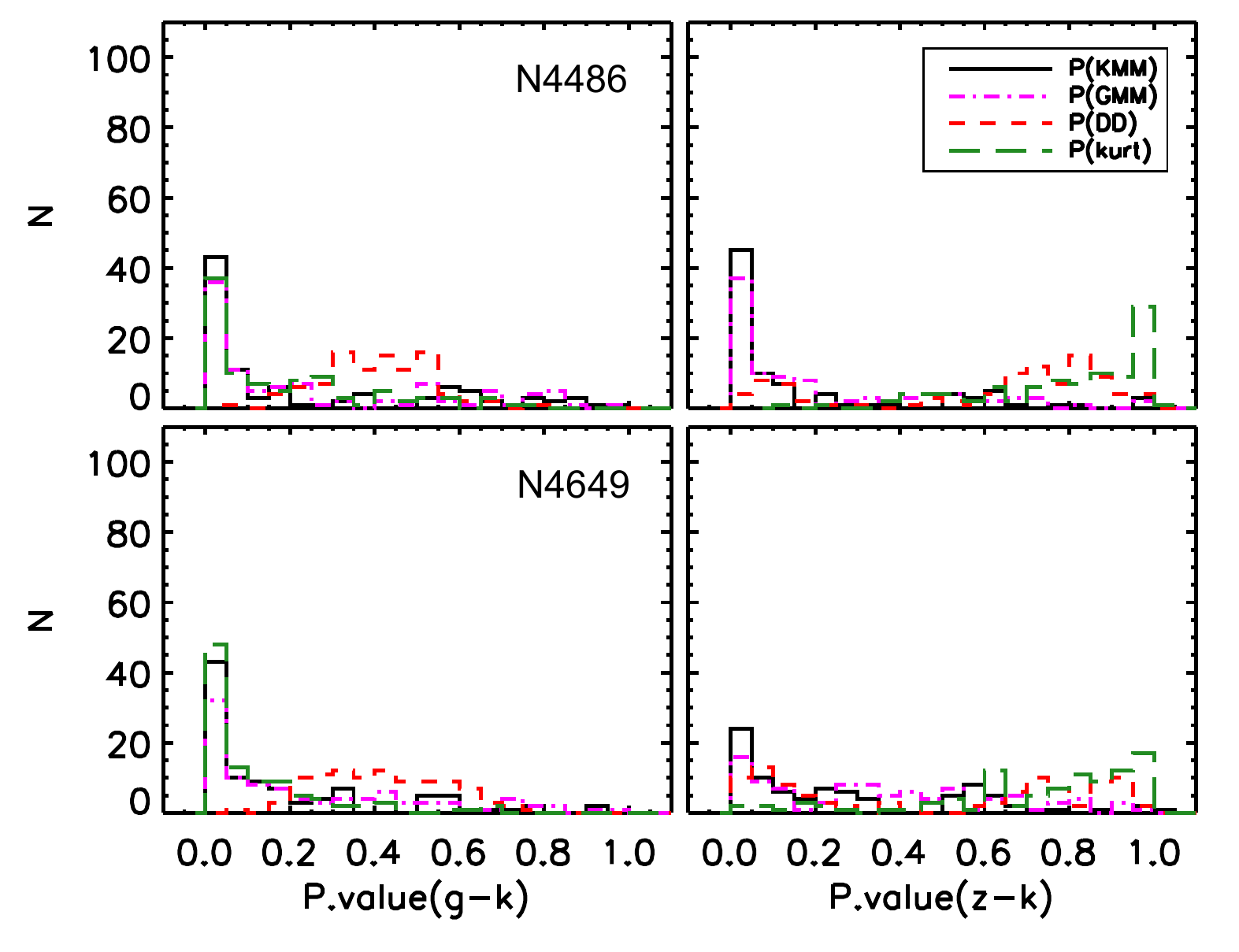}
\caption{Histograms of the probability values returned by KMM and GMM on the output of the simulations for NGC\,4486 (\textit{top panels}) and NGC\,4649 (\textit{bottom panels}) for $(g-K)$ and $(z-K)$.}
\label{outpval_simulsNov11}
\end{figure}

Similarly to the first set, we ran KMM and GMM on the output
  of these set of simulations.  Moreover, before running KMM and GMM we applied the
  same colour cuts ($1<(g-K)<5$ and $0.5<(z-K)<4$) as for the real
  data. In Fig.\,\ref{outpval_simulsNov11} we show the outputs of KMM
  and GMM of the third set of simulations, similarly to Fig.\ref{outpval_simuls}.
  For the third set of simulations these tests converged for $ \gtrsim96$\% of the cases for both KMM and GMM
  for both simulated colours. 
 The $(z-K)$ simulated
  distributions of NGC\,4486 converged for $\sim88$\% of the cases.
  Fig.\,\ref{outpval_simulsNov11} only shows the cases which
  converged. 
  From the left panels of Fig.\,\ref{outpval_simulsNov11} it clear that there 
  is a peak at 0.05 for P(KMM), P(GMM) and P(kurt).
  This indicated that for the $(g-K)$ simulations of NGC\,4486 and NGC\,4649 bimodality is favoured over unimodaliy.
  Formally, $\sim$44\% of the simulations are found to be significantly bimodal considering
  P(KMM) for these GC systems. If one considers P(GMM) these values
  are 38\% for NGC\,4486 and 32\% for NGC\,4649. 
  
  However, the interpretation of the right panels of Fig.\,\ref{outpval_simulsNov11} is slightly more tricky.
  First, the P(kurt) values are skewed to 1 especially for NGC\,4486, which attests the strongly peaked unimodal distributions.
  This, combined with tails of outliers and/or short baseline, will cause KMM and GMM to assign a high probability of bimodality to many of the simulated distributions.
  In fact, KMM (GMM) assigns a high probability of bimodality ($\lesssim 0.05$) for
  $\sim$52\%(27\%) and $\sim$42\%(17\%) of the simulations
  respectively. 
  KMM and GMM assign the great majority of the
  objects to one of the peaks and still give high probability of
  bimodality for several of the simulated $(z-K)$ distributions.
  However, the parent $(g-z)$ distributions show a nearly equal division between blue and red clusters.   

In summary, for the $(g-K)$ distributions of NGC\,4486 and
  NGC\,4649 it is reasonable to say that a bimodal $(g-z)$
  distribution when transformed to $(g-K)$ will remain
  bimodal for a little less than half of the cases and at the same
  time reproduce the observed scatter in $(g-z)$ \textit{vs.} $(g-K)$.
 However, one caveat to keep in mind is that the simulations based on the two colour diagrams are a worst-case scenario that assumes all the additional scatter comes from $(g-K)$.
  The estimate as to how likely it is for a bimodal distribution to
  remain bimodal in the optical/NIR colours in the presence of a
  scatter that reproduces the observed spread in the colour-colour
  diagrams is uncertain for $(z-K)$. We do not show the results of the second set
  of simulations as they are very similar to the third dataset.

One might ask how well do the positions of the peaks of the $(g-K)$ and $(z-K)$ distributions agree with the $(g-z)$ ones when "transformed back" to $(g-z)$ using relations (1) to (6).
For example, for NGC\,4486 and NGC\,4649, using the GMM values of Table\,2 we find that the differences between the "transformed back" and the real distributions is very small, of the order of $\sim0.01$ for relation (1), involving $(g-K)$ and $(g-z)$; but can be as high as $\sim0.2$ for relation (2) involving $(z-K) - (g-z)$.  
When the broken fit relations (3), (4), (5) and (6) are used the differences are $\lesssim 0.06$, being always higher for the relations involving $(z-K)$.
 
Given that $(g-K)$ is a good metallicity proxy, if NGC\,4486 has a unimodal $(g-K)$ distribution, this analysis suggests an underlying unimodal [Fe/H] distribution in the absence of no scatter in the colour-colour diagrams. This is contrary to NGC\,4649, whose $(g-K)$ colour and consequently its underlying [Fe/H], appear genuinely bimodal. 
If colour bimodality is a consequence of metallicity bimodality as typically assumed, then one would have expected more prominent optical/NIR bimodal distributions compared to the optical colour distributions, because the former trace metallicity better than the latter. However, bimodality may be blurred by scatter and is detected in only half of the realizations.

\section{Discussion}
This study shows the appearance of the optical/NIR colour
distributions of GC systems for over a dozen galaxies for the first
time.  The bimodality feature, clear in purely optical colours, is
found to be less pronounced in redder colours. For the most GC-rich
galaxies, NGC\,4486 and NGC\,4649, the K-band photometric scatter does
not appear to be responsible for the almost disappearance of this,
once well-established feature of GC systems. In the absence
  of any scatter in colour-colour diagrams such as $(g-z)$
  \textit{vs.} $(g-K)$ the non linearities caused by the HB
morphology might indeed be the cause of the optical colour bimodality
as suggested by \cite{yoon06} at least in some of the GC systems. This
seems to be the case for NGC\,4486, but not for NGC\,4649.
However, before strong claims are made as to whether certain
  galaxies have or have not underlying bimodal distributions, data of
  better quality are necessary.

In some galaxies, blue GCs tend to be redder at brighter magnitudes. This causes the observed \textit{blue-tilt} (e.g. \citealt{strader06}, \citealt{mieske06}, \citealt{harris06}) feature in the colour magnitude diagrams of certain GC systems. One might argue that the \textit{blue-tilt} combined with the higher photometric scatter present in the optical/NIR colours in comparison to the purely optical colour could be in part responsible for the disappearance of a clear bimodality feature in optical/NIR colours of GC systems such as NGC\,4486 (known to contain the \textit{blue-tilt)}. We argue that there is no reason to think so.
By looking at the NGC\,4486 histograms in Fig.\,\ref{clustrichcoldisits} for clusters with $K_{err}<0.05$, containing only the brightest GCs of the sample, one can see that the $(g-z)$ distribution shows a very clear bimodal feature while the optical/NIR distributions (especially $(z-K)$) do not show it as clearly.  
If the \textit{blue-tilt} were to be (in part) responsible for the non-clear bimodal feature in the optical/NIR colours it would have to also be in the optical one.
Moreover, both \textit{blue-tilt} and bimodality have been detected using $(g-z)$ data.
Furthermore this feature cannot be the main factor behind the absence of a clear $(g-K)$ and $(z-K)$ bimodality in NGC\,4486 compared to NGC\,4649.
The GC systems of both NGC\,4486 and NGC\,4649 have a clear \textit{blue-tilt} (\citealt{strader06}, \citealt{mieske06}).

\cite{cohen98} presented a spectroscopic sample of GCs in NGC\,4486. For a marginal preference of bimodality to be detected at the 89\% significance level, they have to exclude a tail of very metal rich-GCs. Their [Fe/H] distribution is very narrow, in fact much more than that of the Milky Way, as shown in their Fig. 20. 

\begin{figure}
\begin{center}
\includegraphics[width=7cm]{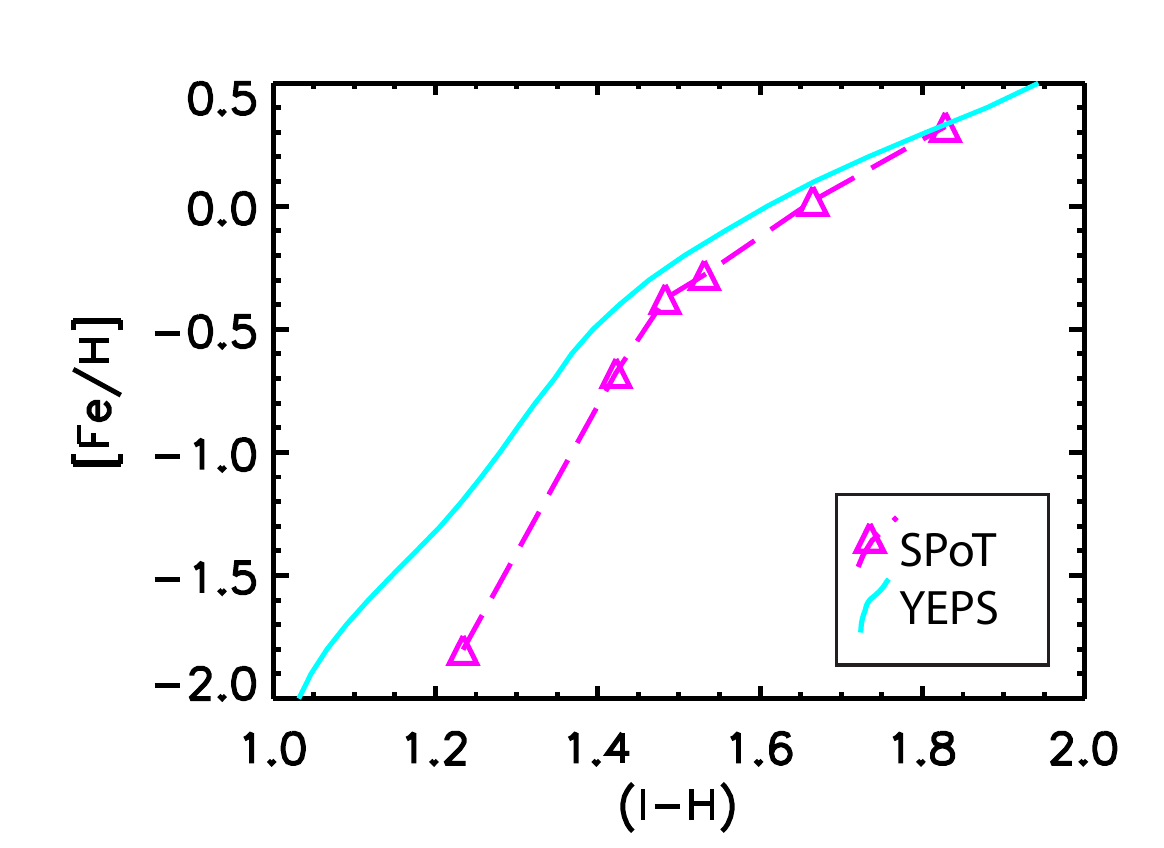}
 \caption{SPoT and Yonsei 14\,Gyr $(I-H)$ - metallicity relation.}
 \label{ihcolmet}
\end{center}
\end{figure}

The fact that \cite{kundu07} found a clear $(I-H)$ bimodal distribution for NGC\,4486 contrasts with our result for unimodality in $(g-K)$.
However, the $(I-H)$ $-$ metallicity relation, shown in Fig. \ref{ihcolmet} for SPoT and YEPS models show a wavy feature that is responsible for projecting equidistant metallicity intervals into larger colour bins.
Could this be argued as the cause for the bimodal $(I-H)$ distribution found by \cite{kundu07}? Note that the $(I-H)$ dip in the colour distribution of NGC\,4486 (Fig. 1 of \citealt{kundu07}) occurs between $1.45-1.65$, exactly where the wavy feature predicted by the SPoT models is.
As shown in Sect 3, the SPoT $(g-K)$-metallicity relation is far more linear in this regime.
Moreover, the field of view of NICMOS, used in \cite{kundu07} is $\sim$4 times smaller than that of LIRIS and ACS. 
Also, the number of clusters shown in the $(I-H)$ distribution of \cite{kundu07} is roughly only 1/4 of the number in this study and much more centrally concentrated, better sampling the red than the blue sub-population. Therefore the sample LIRIS/ACS sample is significantly different from the NICMOS one.

\begin{figure}
\begin{center}
\includegraphics[width=7cm]{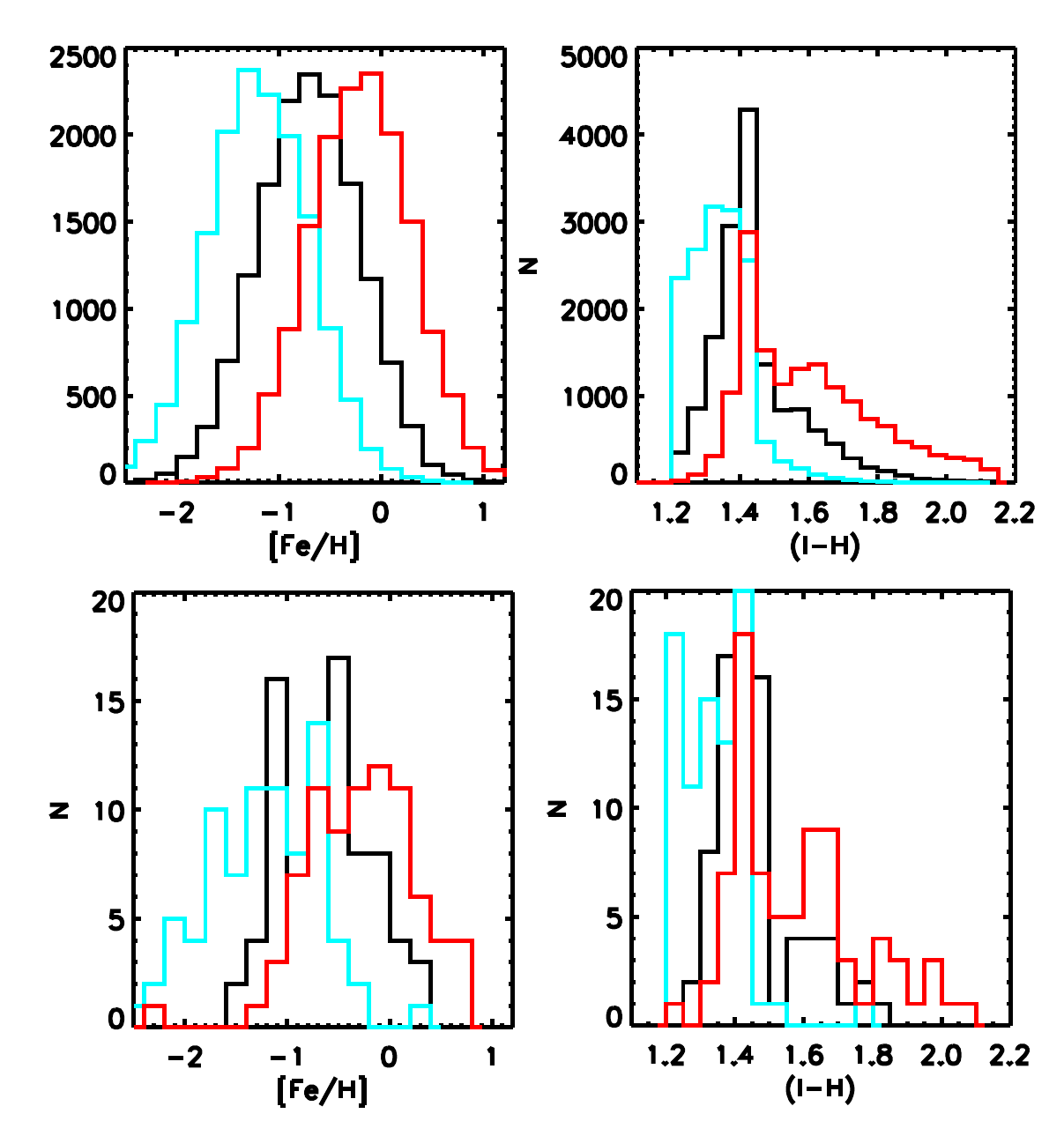}
 \caption{Simulations of unimodal metallicity distributions with different means for 15000 (\textit{upper left panel}) and 80 GCs (\textit{bottom left panel}) transformed to $(I-H)$ according to the 14\,Gyr SPoT  - $(I-H)$ - metallicity relation. The resulting respective $(I-H)$ distributions are shown in the top and bottom right panels.}
 \label{ihcolsimul}
\end{center}
\end{figure}

To further investigate what could cause the bimodality seen in \cite{kundu07} we performed some simulations of unimodal metallicity distributions and transformed them to $(I-H)$ using the 14\,Gyr SPoT model. This is done in order to show that the $(I-H)$ distribution might arise due to this specific colour-metallicity relation combined with the fact the sample is biased to metal-rich GCs.
In the top panel of Fig.\,\ref{ihcolsimul}, examples of these simulations of unimodal metallicity distributions with different means for 15000 (estimated NGC\,4486 total number of GCs (\citealt{peng06})) and 80 (number of GCs in the \citealt{kundu07} sample) GCs are shown. 
They are different gaussian distributions with the same dispersion$\,=\,0.5$ and means$\,=\,0.2$, 0.7 and 1.2. 
The resulting respective colour distributions for these representative examples are shown in the bottom panel of Fig. \ref{ihcolsimul}. Note that the most metal-rich unimodal metallicity distribution becomes bimodal in $(I-H)$ both for the 15000 and for the 80 GCs examples. Note also that the dips occur at the same $(I-H)$ values ($\sim1.5$) of the distribution of \cite{kundu07}. While the simulation that is more closely related to the NGC\,4486
observed metallicity distribution (\citealt{cohen98}) is the intermediate metallicity one, shown in black, the one shown in red matches better the observed data of \cite{kundu07}.
This is probably because the $(I-H)$ observed distribution is biased to more metal-rich values. 
It is well known that GCs in the centres of galaxies are more metal-rich than in their outskirts (e.g. \citealt{liu11}).
One might argue that the $(I-H)$ transformations shown in Fig.\,\ref{ihcolsimul} are idealised as they do not contain photometric scatter. The uncertainty in $(I-H)$ for the \cite{kundu07} study is found to be less than 0.1. Following this, a 0.1 gaussian scatter was added to the transformed $(I-H)$ distributions. 
For the most metal-rich distributions (red histograms), bimodality continues to be seen in $(I-H)$ for the great majority of the cases (with and without the addition of a 0.1 gaussian scatter).

The near universality of GC colour bimodality, once well consolidated as direct implication for metallicity bimodality appears to go contrary to the evidence presented here. In the absence of any scatter in the colour-colour diagrams our analysis shows that the equivalence of bimodal colour and bimodal metallicity distribution is questionable and there is room for some speculation.
Not all luminous galaxies necessarily have bimodal metallicity distributions.
Each galaxy is particular and full of peculiarities, and its metallicity distribution should account for that. If in the Milky Way [Fe/H] bimodality is absolutely evident, this shows that in our Galaxy the major star formation episodes occurred perhaps in less complicated ways. It might be that NGC\,4649 due to its obvious optical/NIR colour bimodality had a similar history to that of the Milky Way.
CD galaxies, such as NGC\,4486 and NGC\,1399 (see the unimodal $(I-H)$ distribution for NGC\,1399 found by \citealt{blakeslee12}), are much more prone to interactions with their neighbours due to their physical position in the potential well of the galaxy cluster. Perhaps the quantity and/or strength of such interactions play an important role in determining the shape of the [Fe/H] distributions observed today. This picture of non-universality of [Fe/H] bimodality fits much better with the hierarchical merging paradigm. 

High S/N NIR/optical imaging in an 8-10m class telescope is of the utter most importance to further investigate this issue. Also, large spectroscopic datasets coupled with simulations for several GC systems would shed more light in the true nature of the metallicity distributions of GC systems in large galaxies.

\section{Summary and conclusions}
We have analysed the $(g-z)$, $(g-K)$ and $(z-K)$ colour distributions of GC systems in a sample of 14 E/S0 galaxies.
The results are summarised below:
\begin{enumerate}
\item The data presents a non-linear feature around $(z-K)\,\sim\,2$ and $(g-K)\,\sim\,3.2$, marking the transition from blue to red HB morphology. SPoT and YEPS models, with a realistic treatment of the HB also show a similar feature. 
According to these models for old ages the colour metallicity relation does not present the prominent wiggle of \cite{yoon06} for the optical/NIR colours: $(g-K)$ and $(z-K)$.
\item While the great majority of GC systems present an obvious bimodal 
distribution in $(g-z)$, bimodality is clearly less pronounced in the 
optical/NIR distributions $(g-K)$ and $(z-K)$. 
The two most cluster rich galaxies show some remarkable differences. While
the GC system of NGC\,4486 shows no obvious bimodality 
in $(g-K)$ when all the GC sample is considered, the GC system of NGC\,4649 does.
However, if restricted to a brighter sub-sample with small K-band errors ($<$ 0.05 mag) the $(g-K)$ distribution of NGC\,4486 GCs is better described by a bimodal distribution.
Simulations of the $(g-K)$ and $(z-K)$
distributions with realistic K-band errors suggest that the K-band errors 
cannot be the responsible for the blurring of genuinely bimodal 
optical/NIR colour distributions in NGC\,4486 and NGC\,4649. 
However, when taking into account the extra scatter present in colour-colour diagrams such as $(g-K)$ \textit{vs.} $(g-z)$, we find that bimodality
is indeed likely to be undetectable in over half of the cases for the 
$(g-K)$ distributions of these two GC systems.
\item 
The underlying metallicity distribution of the GC system of NGC\,4649 appears 
to be a genuine case for bimodality. However, for NGC\,4486 the situation
is less clear and bimodality is detected at a statistically significant
level only for the brightest sub-sample of the clusters in $(g-K)$. This bimodality becomes less pronounced
when including objects with larger errors, or for the $(z-K)$ colour
distributions. 
In the galaxy, centre of the Virgo Cluster, the argument put forward by \cite{yoon06} might contribute for the clear \textit{optical} colour bimodality.
Higher S/N NIR imaging is strongly needed to understand whether the $(g-K)$ lack of bimodality for NGC\,4486 could be due to scatter other than photometric. Also it would further constrain the nature of the optical/NIR colour distributions (and the underlying metallicity distributions) of the other, less GC-rich galaxies in the sample.

\end{enumerate}

\begin{acknowledgements}
We thank Oleg Gnedin for providing us with the GMM algorithm and the anonymous referee for putting a lot of effort in reviewing this manuscript. His suggestions, significantly improved the paper.
\end{acknowledgements}


\begin{thebibliography}{}

\bibitem[Alves-Brito et al.(2011)]{alves11} Alves-Brito, A., 
Hau, G.~K.~T., Forbes, D.~A., et al.\ 2011, \mnras, 417, 1823 
\bibitem[Ashman \& Zepf(1992)]{az92}Ashman, K. M. \& Zepf, S. E., 1992, ApJ, 384, 50
\bibitem[Ashman \& Zepf(1993)]{az93}Ashman, K. M. \& Zepf, S. E., 1993 MNRAS, 264, 611
\bibitem[Ashman, Bird \& Zepf (1994)]{abz94} Ashman, K.~M., Bird, 
C.~M., \& Zepf, S.~E.\ 1994, \aj, 108, 2348 
\bibitem[Beasley et al.(2002)]{beasley02}Beasley, M.A., Baugh, C. M., Forbes, D. A., Sharples, R. M., Frenk, C. S. , 2002, MNRAS, 333, 383
\bibitem[Beasley et al.(2008)]{beasley08}Beasley, Michael A., Bridges, T., Peng, E. et al., 2008, MNRAS, 386, 1443
\bibitem[Blakeslee et al.(2010)]{blakeslee10} Blakeslee, J.~P., Cantiello, M., \& Peng, E.~W.\ 2010, \apj, 710, 51 
\bibitem[Blakeslee et al.(2012)]{blakeslee12} Blakeslee, J.~P., 
Cho, H., Peng, E.~W., et al.\ 2012, arXiv:1201.1031 
\bibitem[Bica et al.(2006)]{bica06} Bica, E., Bonatto, C., Barbuy, B., \& Ortolani, S.\ 2006, A\&A, 450, 105 
\bibitem[Bird et al.(2010)]{bird10} Bird, S., Harris, W.~E., Blakeslee, J.~P., \& Flynn, C.\ 2010, \aap, 524, A71 
\bibitem[Biscardi et al.(2008)]{biscardi08} Biscardi, I., 
Raimondo, G., Cantiello, M., \& Brocato, E.\ 2008, \apj, 678, 168 
\bibitem[Brodie \& Strader(2006)]{bs06} Brodie J.P. \& Strader J. 2006, ARA\&A, 44, 193
\bibitem[Brocato et al.(2000)]{brocato00} Brocato, E., Castellani, V., Poli, F.~M., \& Raimondo, G.\ 2000, A\&AS, 146, 91 
\bibitem[Cantiello \& Blakeslee(2007)]{cb07} Cantiello, M. \& Blakeslee, J.P., 2007, ApJ, 669, 982
\bibitem[Caldwell et al.(2011)]{caldwell11} Caldwell, N., 
Schiavon, R., Morrison, H., Rose, J.~A., \& Harding, P.\ 2011, AJ, 141, 61 
\bibitem[Chies-Santos et al.(2011a)]{paper1} Chies-Santos, A. L., Larsen, S. S., Wehner, E. M., Kuntschner, H., Strader, J., \& Brodie, J.~P.\ 2011, A\&A, 525, A19 
\bibitem[Chies-Santos et al.(2011b)]{paper2} Chies-Santos, A. L., Larsen, S. S., Kuntschner, H., Anders, P., Wehner, E. M., Strader, J., Brodie, J.~P., \& Santos, J. F. C.\ 2011, A\&A, 525, A20 
\bibitem[Cohen et al.(1998)]{cohen98} Cohen, J. G., Blakeslee, J. P. \& Ryzhov, A., 1998, ApJ, 496, 808
\bibitem[Cohen et al.(2003)]{cohen03} Cohen, J.~G., Blakeslee, 
J.~P., \& C{\^o}t{\'e}, P.\ 2003, \apj, 592, 866 
\bibitem[C\^ot\'e et al.(1998)]{cote98} C\^ot\'e, P., Marzke, R.O., West, M.J., 1998, ApJ, 501, 554
\bibitem[Dirsch et al.(2003)]{d03}Dirsch, B., Richtler, T., Geisler, D. et al. 2003, AJ, 125, 1908
\bibitem[Elson \& Santiago(1996)]{es96}Elson, R. A. W. \& Santiago, B. X., 1996, MNRAS, 278, 617
\bibitem[Forbes et al.(1997)]{fbg97}Forbes, D.A., Brodie, J. P., Grillmair, C. J., 1997, AJ, 113, 1652
\bibitem[Foster et al.(2010)]{foster10} Foster, C., Forbes, 
D.~A., Proctor, R.~N., Strader, J., Brodie, J.~P., 
\& Spitler, L.~R.\ 2010, \aj, 139, 1566 
\bibitem[Foster et al.(2011)]{foster11} Foster, C., 2011, MNRAS, 415, 3393
\bibitem[Harris et al.(2006)]{harris06} Harris, W.~E., Whitmore, B.~C., Karakla, D., Oko{\'n}, W., Baum, W.~A., Hanes, D.~A., \& Kavelaars, J.~J.\ 2006, ApJ, 636, 90 
\bibitem[Koekemoer et al.(2002)]{koek02}Koekemoer, A.~M., 
Fruchter, A.~S., Hook, R.~N., 
\& Hack, W.\ 2002, The 2002 HST Calibration Workshop : Hubble after the Installation of the ACS and the NICMOS Cooling System, 337 
\bibitem[Kundu \& Zepf(2007)]{kundu07}Kundu, A. \& Zepf, S. E., 2007, ApJ, 660,190
\bibitem[Larsen(1999)]{larsen99}Larsen,S.S.,1999, A\&AS, 139, 393
\bibitem[Lee et al.(1994)]{lee94}Y. W. Lee, P. Demarque, R. Zinn, 1994, ApJ, 423, 248
\bibitem[Liu et al.(2011)]{liu11} Liu, C., Peng, E.~W., Jord{\'a}n, A., Ferrarese, L., Blakeslee, J.~P., C{\^o}t{\'e}, P., \& Mei, S.\ 2011, ApJ, 728, 116 
\bibitem[Maraston(2005)]{m05}Maraston, C., 2005, MNRAS, 362, 799
\bibitem[Mieske et al.(2006)]{mieske06}Mieske, S., Jord\'an, A., C\^ot\'e,P. et al., 2006, ApJ, 653, 193
\bibitem[Marigo et al.(2008)]{marigo08}Marigo, P., Girardi, L., Bressan, A. et al., 2008, A\&A, 482, 883
\bibitem[Muratov \& Gnedin(2010)]{muratov10} Muratov, A.~L., \& Gnedin, O.~Y.\ 2010, ApJ, 718, 1266 
\bibitem[Peng et al.(2006)]{peng06}Peng, E. W,  Jord\'an, A.,  C\^ot\'e, P. et al., 2006, ApJ, 639, 95
\bibitem[Peng et al.(2008)]{peng08}Peng, E. W,  Jord\'an, A.,  C\^ot\'e, P. et al., 2008, ApJ, 681, 197
\bibitem[Woodley et al. (2010)]{woodley10}Woodley, K. A., Harris, W. E., Puzia, T. H., et al., 2010, ApJ, 708, 1335
\bibitem[Raimondo et al.(2005)]{raimondo05}Raimondo, G., Brocato, E., Cantiello, M. \& Capaccioli, M., 2005, AJ, 130, 2625 
\bibitem[Renzini(2006)]{renzini06} Renzini, A., 2006 ARA\&A, 44, 141
\bibitem[Richtler(2006)]{richtler06} Richtler, T.\ 2006, Bulletin 
of the Astronomical Society of India, 34, 83 
\bibitem[Rhode et al.(2005)]{rhode05}Rhode, K.L., Zepf, S.E., Santos,R., 2005, ApJ, 630, 21
\bibitem[Sohn et al.(2006)]{sohn06} Sohn, S.~T., O'Connell, 
R.~W., Kundu, A., Landsman, W.~B., Burstein, D., Bohlin, R.~C., Frogel, 
J.~A., \& Rose, J.~A.\ 2006, \aj, 131, 866
\bibitem[Spitler et al.(2008)]{spitler08} Spitler, L.~R., Forbes, D.~A., \& Beasley, M.~A.\ 2008, MNRAS, 389, 1150 
\bibitem[Stetson(1987)]{stetson87}Stetson, P. B., 1987, PASP, 99, 191
\bibitem[Strader et al.(2005)]{strader05}Strader,J., Brodie, J.P., Cenarro,A.J.,Beasley,M.A. \& Forbes, D.A. 2005, AJ,130,1315
\bibitem[Strader et al.(2006)]{strader06}Strader,J., Brodie, J.P., Spitler, L., Beasley, M. A., 2006, AJ, 132, 2333  
\bibitem[Strader et al.(2007)]{strader07} Strader, J., Beasley, 
M.~A., \& Brodie, J.~P.\ 2007, \aj, 133, 2015
\bibitem[Yoon et al.(2006)]{yoon06}Yoon, S.J. and Yi, S.K. and Lee, Y.-W., 2006, Sci, 311,1129
\bibitem[Yoon et al.(2011a)]{yoon11a} Yoon, S.-J., Sohn, S.~T., 
Lee, S.-Y., et al.\ 2011, \apj, 743, 149
\bibitem[Yoon et al.(2011b)]{yoon11b} Yoon, S.-J., Lee, S.-Y., 
Blakeslee, J.~P., et al.\ 2011, \apj, 743, 150 
\bibitem[Zinn(1985)]{zinn85} Zinn, R.\ 1985, ApJ, 293, 424 


\end{thebibliography}
\end{document}